\documentstyle[prl,floats,aps,twocolumn,multicol,epsf]{revtex}
\tighten
\begin{document}
\twocolumn[\hsize\textwidth\columnwidth\hsize\csname
@twocolumnfalse\endcsname
%\baselineskip 28 pt plus2pt
%***************************************************************************
%**************************************************************************
%**************************************************************************
\title{Interaction of electromagnetic perturbations with infalling
observers inside
spherical charged black holes}
\author{Lior M. Burko}
\address{Department of Physics,
Technion---Israel Institute of Technology,32000 Haifa, Israel}
\date{\today}
\maketitle 

\begin{abstract}
The electromagnetic radiation that falls into a Reissner-Nordstr\"{o}m
black hole is known to develop 
a ``blue sheet'', namely, an infinite concentration of energy density at
the Cauchy horizon. The interaction of these divergent electromagnetic
fields with infalling matter was recently analyzed (L. M. burko and A.
Ori, Phys. Rev. Lett. {\bf 74}, 1064 (1995)). Here, we give a more
detailed 
description of that analysis: We consider classical electromagnetic fields
(that were
produced during the collapse and then backscattered into the black hole),
and investigate the blue-sheet effects of these fields on infalling objects
within two simplified models of a classical and a quantum absorber. 
These effects are found to be finite and even 
negligible for typical parameters of a supermassive black hole.
\end{abstract}

\vspace{3ex}
]

\section{Introduction}
The Kerr-Newman black hole (BH), which in view of the no-hair theorems is 
expected to be the stationary outcome of gravitational collapse, is an exact 
(electro-) vacuum solution of the Einstein field equations. 
However, one expects a 
generic collapse process to be 
accompanied by perturbations, which may exist before 
the onset of the collapse, or develop during it. 
Consequently, one would not expect 
to find an astrophysical BH to be an {\em exact} 
Kerr-Newman solution, but rather 
a perturbed one.  

The possibility to fall into a black hole and 
re-emerge in another universe is one of
the most intriguing open questions of General Relativity \cite{novikov}.
The spacetime of unperturbed
BHs, such as Kerr, seems to allow this possibility. However,
the Kerr geometry is highly
symmetrical; It is not {\em a priori} clear whether more realistic
solutions to the Einstein
equations, which are not unperturbed, still allow for the possibility to
traverse the inner horizon safely and re-emerge in a different universe. 
Consequently, perturbed BHs have been under investigation
during the last three decades.

Realistic BHs are the outcome of
gravitational collapse. In such a realistic (a-symmetric) collapse,
non-vanishing multipole moments of various fields develop in the star,
and consequently 
electromagnetic and gravitational waves are emitted from the surface of the
collapsing star. As these waves propagate outwards, some fraction of them
is backscattered off the spacetime curvature and captured by the BH. This
process leads to a ``tail" of radiation,
which at late times decays according to an inverse-power law both for the 
spherical case \cite{price,gundlach,bicak,burko-ori97} and for the 
spinning case \cite{krivan1,krivan2,barack}. 

To re-emerge in another universe it is
necessary to traverse a certain null hypersurface known as the inner horizon
or the Cauchy horizon (CH). In order to cross the CH safely,
it is necessary that the spacetime be
either non-singular there or have at the most only a weak singularity.  
Therefore, it is troubling that the CH is a surface of
infinite blue-shift, i.e., infalling radiation (electromagnetic or
gravitational), even very mild and 
well-behaved 
in the external universe, is infinitely blue-shifted at the
CH \cite{penrose}. 
There are two types of problems which may cause difficulties for an
observer who wishes to cross the CH. First, a divergent spacetime 
curvature develops on the CH. This curvature singularity 
results from two sources: the 
infinite blue shift of
infalling gravitational waves (which lead to the divergence of the
gradients of the metric perturbations) 
and the infinite energy-momentum tensor (which can be taken as a 
dynamic 
source term for the Einstein equations) of the infinitely blue-shifted
infalling electromagnetic radiation. Second, the same infinitely
blue-shifted electromagnetic waves cause an infinite flux of radiation, 
which might heat any infalling observer unlimitedly. 

Penrose \cite{penrose} argued that the CH was unstable against small
perturbations. His arguments for the infinite blue shift at the CH
relied on a geometric-optics approximation. Penrose argued that
perturbations originated in an infinitely long (external) time are
concentrated in a finite (proper) time of the infalling object. Thus,
unless these perturbations decay at least exponentially fast in external
time, the infinite concentration will lead to the blue sheet. Later,
other works considered the wave equation for the evolving perturbations.
Simpson and Penrose \cite{simpson}
re-affirmed numerically (for linear electromagnetic
perturbations) the qualitative
arguments of Penrose. G\"{u}rsel {\it et al}
\cite{gursel} and Chandrasekhar and Hartle \cite{chandrasekhar}
calculated the projection of the energy-momentum tensor on the world-line 
of an infalling observer, and inferred from its divergence on the CH that 
the radiation absorbed by the observer was also divergent. Nevertheless,
the {\it fundamental} fields (i.e., the scalar field, the
electromagnetic four-potential, and the metric perturbations) were found to
be regular. It is the gradients of the fundamental fields which diverge on
the CH. Therefore, it was suggested by Ori \cite{ori1,ori2}
that although there is a true
curvature singularity at the CH, this singularity is rather weak. Namely,
despite the divergence of the tidal force, the actual
tidal distortion experienced by infalling observers (as they hit the
singular CH) is finite, and for
typical parameters  -- even negligible. To cross the CH safely there still
remains, though, the other
aforementioned potential problem. Namely, the possible complete burning up
of any physical body at the CH, due to the divergent electromagnetic field
there (and the associated energy flux). This subject was considered
recently by Burko and Ori \cite{burko}. In this Paper we give a more
detailed account of that work.

Physical BHs are expected to spin
very fast. However, it 
turns out that the mathematical analysis involved 
with the evolution of the multipole moments is very complicated, due to
the axial symmetry of the Kerr background.
For this reason, it is often common to work with a toy model, within which
the mathematical analysis is much simpler. Of course, the toy model should
preserve the most essential and relevant properties of the realistic
BH. Thus, most work (including the present one) is done in the framework of the
Reissner-Nordstr\"{o}m (RN) 
BH. (The RN solution is involved, however, with a 
certain complication, which arrises from the coupling of the gravitational 
and the electromagnetic fields.)  
The RN spacetime is the unique electrically charged, 
spherically symmetric, static vacuum solution of the coupled 
Einstein-Maxwell 
equations. For obvious reasons physical BHs are not
expected to be significantly charged. The vanishing angular momentum of
the RN BH is also an unrealistic feature. 
Nevertheless, it turns out that the internal causal structure of the RN BH
is very similar to the internal structure of the Kerr BH: 
In both spacetimes the singularities are
timelike, and are located beyond a CH;
Both spacetimes have a wormhole-like topology, which may allow for a travel
to other asymptotically-flat universes. Consequently, it is
believed that the RN BH is a physically justifiable model for realistic
BHs.

In this paper we analyze the interaction of an infalling object with the 
divergent electromagnetic field which we expect to find at the CH. 
Throughout this paper we shall assume 
that the infalling object is much smaller than the typical
radius of curvature near the CH. 
However, electromagnetic perturbations in the RN geometry are always
accompanied by gravitational perturbations. Typically, the latter produce
a diverging curvature at the CH (due to the infinite blue shift), which
means that the radius of curvature vanishes there. 
Our assumption is valid only if---for the sake of evaluating 
the radiative electromagnetic effects---we
ignore the gravitational perturbations. The modification of the
radiative interaction by the metric perturbations is obviously a non-linear
effect, as it is quadratic in the perturbation's amplitude. This non-linear 
effect remains the subject of future research. We do not expect, however, 
this non-linear effect to significantly alter the linear-order interaction. 
The gravitational analogue of this problem---the object's interaction with 
the divergent tidal forces---demonstrates this reasoning: As implied from the
analysis of Ref. \cite{ori2} on the strength of the CH singularity in 
spinning BHs---where it has been demonstrated
that the non-linear gravitational interaction with an object
may be negligible---we do not expect
higher-order contributions to change the general picture significantly.
Thus, in this work we restrict ourselves to linear
effects only, and take the background to be RN (and not a
gravitationally-perturbed background). In addition, recent
fully-nonlinear numerical simulations have shown that at the asymptotic
past of the CH the metric perturbations vanish, in accord with
perturbation analyses \cite{burko97}. 

The organization of this Paper is as follows:
In section II we summarize the definitions and the notation we use.
In section III we briefly review the formalism given by Chandrasekhar
\cite{chandrasekhar2} for the
determination of the tetrad components of the electromagnetic field
tensor $F_{\alpha\beta}$ at the CH for given perturbations.
In section IV we obtain the
tetrad components of the electric and magnetic fields 
for initial perturbations which decay
according to an inverse-power law. In section V we transform these tetrad
components to their tensorial counterparts and write them in the rest-frame 
of a freely-falling observer.
In section VI and section VII we use these
fields to calculate their interaction with (very simplified) classical and
quantum absorbers, respectively. We then discuss 
(section VIII) the results, and argue 
that although the electromagnetic fields diverge at the CH, the interaction
of the field does not necessarily cause ultimate destruction of the
infalling observer. 

It should be noted that we only treat here classical
radiation, and ignore quantum processes, especially pair-production 
effects.
When these effects are taken into consideration 
\cite{burko3}, it should be expected that
they may change our results here considerably. Yet, we believe that our
main conclusion, namely, 
that the singular CH is {\em not} the edge of spacetime, will
still be relevant even after consideration of the quantum effects.

\section{Definitions and Notation}
We write the line element of an unperturbed RN BH in the form 
\begin{eqnarray}
\,ds^{2}=
e^{2\nu}\left(\,dx^{0}\right)^{2}-e^{2\mu _{2}}\left(\,dx^{2}\right)^{2}
-r^{2}\,d\Omega^{2}, 
 \label{RN}
\end{eqnarray} 
where the coordinates are  
$\left(x^{0}\;x^{1}\;x^{2}\;x^{3}\right)=\left(t\;\phi\;r\;\theta\right)$, 
$\,d\Omega^{2}$ is the unit two-sphere line element, and the metric 
coefficients are $e^{2\nu}=e^{-2\mu_{2}}=(r^{2}-2Mr+Q_{*}^{2})/r^{2}
\equiv \Delta/r^{2}$, 
$M,Q_{*}$ being the mass and electric charge, respectively, of the 
RN BH, and where $r$ is the radial 
Schwarzschild coordinate, defined such that circles of radius $r$ have 
circumference $2\pi r$. The general form of the line element (1) is 
preserved under polar perturbations (sometimes called even-parity 
perturbations); On the other hand, axial perturbations (called also odd-parity 
perturbations), will lead in general to non-vanishing off-diagonal metric. 
[Axial perturbations are characterized by the 
non-vanishing of the metric functions $\omega, q_{2}, q_{3}$ (the 
non-vanishing of these 
metric coefficients induce a dragging of the inertial-frame 
and impart a rotation to the BH), while polar 
perturbations are those which alter the values of the metric functions 
$\nu, \mu_{2}, \mu_{3}$ and $\psi$ (which are in general non-zero for the 
unperturbed BH).] Therefore, the form of the metric of a 
generally-perturbed RN BH  
will be more complicated than the line element (1). 
It has been shown \cite{chandrasekhar2}, 
that a metric of sufficient generality is of the form  
\begin{eqnarray}
\,ds^{2}&=&e^{2\nu}\left(\,dx^{0}\right)^{2}-e^{2\psi}\left(\,dx^{1}-\omega
\,dx^{0}-q_{2}\,dx^{2}\right.\nonumber\\
&-&\left. q_{3}\,dx^{3}\right)^{2} 
-e^{2\mu_{2}}\left(\,dx^{2}\right)^{2}-e^{2\mu_{3}}
\left(\,dx^{3}\right)^{2}.
\end{eqnarray}
Since the unperturbed RN background is spherically 
symmetric, we may consider only axisymmetric perturbation modes without 
any loss of generality. (This is because all non-axisymmetric modes can 
be received from the axisymmetric modes, if the unperturbed spacetime is 
spherically symmetric \cite{chandrasekhar2}.) The metric (2) involves 
seven  functions, namely, $\nu ,\psi ,\mu_{2} ,\mu_{3} ,\omega ,q_{2},$ and 
$q_{3}$. Beacuse the Einstein equations involve only six independent 
functions, not all of the 
seven functions can be determined independently, and 
there is a gauge-fixing freedom on the metric coefficients. However, this 
gauge freedom involves only the metric coefficients $\omega, q_{2}$, and 
$q_{3}$; There is no gauge freedom in the determination of the polar 
perturbations. 

The horizons of the RN BH are the event horizon 
$r_{+}$ and the inner horizon $r_{-}$, which are located at the roots of 
$\Delta$, namely, at $r_{\pm}=M\pm (M^{2}-Q_{*}^{2})^{1/2}$. We denote the 
surface gravity of the event horizon and the CH by $\kappa_{\pm}\equiv 
(r_{+}-r_{-})/r_{\pm}^{2}$, respectively.   
We define the Eddington-Finkelstein null coordinates 
$u=r_{*}-t$ and $v=r_{*}+t$, where 
$r_{*}$ is the Regge-Wheeler `tortoise' coordinate defined by 
$d/\,dr_{*}=(\Delta/r^{2})d/\,dr$. The coordinate $t$ is 
spacelike between the event and the Cauchy horizons, and we take 
$t=+\infty$ at the event horizon. In this Paper we 
are interested in the sections of the event horizon represented by 
$u=-\infty$ and of the inner horizon represented by 
$v=+\infty$. (These are the sections which 
intersect in the 
standard Penrose diagram at future timelike infinity of the external 
universe.) We assume that the object moves along a typical radial 
world line that intersects the event horizon and the CH at 
some finite values $v=v_{0}$ and $u=u_{0}$, respectively. 
Accordingly, the trajectory of the object can be described by the function 
$r(\tau )$ and by $u_{0}$, where $\tau$ is the proper time of the 
infalling object. We set $\tau(r=r_{-})=0$. 

\section{The Chandrasekhar Formalism}
In this section we briefly review the Chandrasekhar formalism for the 
evolution of the polar perturbations 
\cite{chandrasekhar2}, and describe the algorithm which can be constructed 
from it for the determination of the Maxwell tensor.  
The Chandrasekhar formalism fails for the description of dipole 
perturbations, i.e., for the $l=1$ modes in the multipole 
expansion of the perturbations. However, for the treatment of all modes with 
multipole order higher than the dipole, it can still be used.  
The formalism for dipole polar modes was treated 
by Burko and is given in Ref. \cite{burko1}. The modified formalism of Ref. 
\cite{burko1} can be used in an analogous way for the determination of the 
dipole perturbations. 
 
Let us consider a RN BH perturbed by the electromagnetic waves, which we
expect to exist for any charged BH created by a collapse process.
Let the metric of the unperturbed RN background be given by Eq. (\ref{RN}).
The metric form (\ref{RN}) does not change when the collapse is endowed 
with
polar perturbations \cite{chandrasekhar2}.
[Axial perturbations, on the other
hand, will lead to the appearance of other (off diagonal) 
non-vanishing metric
coefficients, and thus complicate our calculations considerably. For this
reason, and as knowledge of the perturbations of either of the classes
enables us to know the perturbations of the other class (to be explained
below), we consider here polar perturbations only. It should be remembered 
that our
entire analysis is linear in the perturbations, and therefore effects of
non-vanishing off-diagonal metric coefficients are negligible.]
We take---after Price
\cite{price},
who considered the uncharged Schwarzschild background,
and Bi\v{c}\'{a}k \cite{bicak}, who extended the treatment to the RN
background---the 
perturbing fields to decay according to the $(2l+2)$ inverse-power
in external time. The infalling radiation propagates through the curved
spacetime of the BH. Due to the RN background, the (linear) gravitational
and electromagnetic perturbations are coupled. However, their equations can
be decoupled \cite{chandrasekhar2,chandrasekhar3}. 

The metric perturbations, analyzed into
their normal modes with a time-dependence $e^{i\sigma t}$ are governed by
the Moncrief-Zerilli \cite{mon+zer}
equations, which are a pair of decoupled one-dimensional 
wave equations of the form
\begin{eqnarray}
\frac{\,d^{2}Z_{i}^{(\pm )}}{\,dr_{*}^{2}}+\sigma^{2}Z_{i}^{(\pm )}=
V_{i}^{(\pm )}Z_{i}^{(\pm )}.
\end{eqnarray}
Here 
$V_{i}^{(\pm )}=\pm\beta_{i}\,df_{i}/\,dr_{*}+\beta_{i}^{2}f_{i}^{2}
+\kappa f_{i}$, where
$\kappa=\mu^{2}(\mu^{2}+2)$, $\beta_{i}=q_{j}$ and
$f_{i}=\Delta /[r^{3}(\mu^{2}r+q_{j})]$. Here, $i,j=1,2$ and $i\ne j$. 
We also take
$\mu^{2}=(l-1)(l+2)$, $q_{1,2}=3M\pm\sqrt{9M^{2}+4Q_{*}^{2}\mu^{2}}$, 
$l$ being the multipole order of the perturbing radiation. 
The superscript $(+)$ denotes polar
perturbations, and $(-)$ denotes axial perturbations. 
As shown in Ref. \cite{chandrasekhar2}, knowledge of modes of each of the 
parities enables us to find the modes of the other parity, so without any 
expected
loss of generality we
shall restrict ourselves here to the treatment of polar perturbations only.  
(It should be stressed that these expressions are inapplicable for dipole 
modes. For the determination of the latter one must use the formalism of 
Ref. \cite{burko1}.) 

Through the end of this section we shall briefly summarize the algorithm we 
can construct from the Chandrasekhar formalism (or from its generalization 
to dipole modes) for the calculation of the perturbed 
metric functions and the components of the Maxwell field strength tensor, 
given the initial perturbing fields $Z_{i}^{(\pm)}$, $(i=1,2)$.  
When we give refernce to a specific equation in Chapter 5 
of Ref. \cite{chandrasekhar2}, a letter ``C'' will precede the equation 
number. 
We first find the functions $H_{i}^{(+)}$ by the algebraic
equations (C186)-(C187). 
Now, we calculate the function $\Phi$ by Eq. (C196). The
next step is to obtain the functions defined by Eqs. (C190), and complete 
the solution with Eqs.  (C191)-(C195).
The perturbations of the metric $(1)$ will be given then by Eqs. (C158) 
and the perturbations in the tetrad
components of the Maxwell field strength tensor will be given by Eqs. 
(C159-C160). In the following we shall transform from the tetrad components 
of the Maxwell tensor to their tensorial counterparts using the tetrad 
\begin{eqnarray}\left \{ \begin{array}{l}
e_{(0)\mu}=e^{\nu} (1\;0 \; 0 \; 0) \nonumber\\
e_{(1)\mu}=r\sin \theta (0 \; 1 \; 0 \; 0) \nonumber\\
e_{(2)\mu}=e^{-\nu} (0 \; 0 \; 1 \; 0) \nonumber\\
e_{(3)\mu}=r (0 \; 0 \; 0 \; 1)\nonumber .
\end{array} \right. \end{eqnarray}

\section{Derivation of the Maxwell Field Near the CH  
-- Tetrad Components}

In what follows, we consider an isolated charged BH, 
surrounded by electromagnetic waves, which we treat as a linear 
perturbation. (In fact, because of the 
non-vanishing electric field of the background, 
this linear perturbation consists of both electromagnetic 
and gravitational waves \cite{gursel,chandrasekhar}.)  
We calculate the asymptotic behavior of 
the electromagnetic perturbation near the CH. 
The class of perturbations that we consider here is the one which is 
inherent to any non-spherical gravitational-collapse; 
These are the electromagnetic    
perturbations which result from the evolution of non-vanishing 
electromagnetic  
multipole moments (in the star) during the collapse. When these 
perturbations propagate outwards, some fraction of them is 
backscattered off the spacetime curvature and captured by the BH. 
This process leads to a ``tail'' of infalling radiation at the event 
horizon which at late times $(v\gg M)$ decays like $(v/M)^{-(2l+2)}$, 
where $l$ is the multipole order of the mode \cite{price}. We treat this 
electromagnetic field according to the formalism by Chandrasekhar 
\cite{chandrasekhar2} for 
$l\ge 2$ polar modes and the extension of this formalism by Burko 
\cite{burko1} for dipole ($l=1$) polar modes.  (The extension of the formalism 
is needed as we {\it a posteriori} find that the effects we study are 
dominated by the dipole mode. This mode is not treated properly 
by Ref. \cite{chandrasekhar2}.) 

\subsection*{The Components of $B_{\mu\nu}$}
Our goal now is to find an approximate expression [to the leading order in 
$(r-r_{-})$ and $(\kappa_{-}u_{0})^{-1}$] 
for the electromagnetic field an 
infalling observer measures on arrival at the CH. We shall 
find that 
some components of the Maxwell field strength tensor diverge there, while 
other components remain finite. Therefore, we shall restrict ourselves here 
to the evaluation of the divergent components only. (The other 
components
can be found analogously.) We shall consider here an $l\ge 2$ 
polar mode of
infalling radiation. It should be stated that the most dominant effect is
{\it not} caused by the $l\ge 2$ modes but by
the $l=1$ mode. The reason we choose here to discuss the $l\ge 2$ modes
rather than the $l=1$ mode is that the formalism of Ref.
\cite{chandrasekhar2} is inapplicable for the treatment of
dipole modes, as mentioned in section III (see 
Ref. \cite{burko1} for details). It turns out, that when one calculates
the dipole perturbations
according to the generalized formalism of Ref. \cite{burko1} the results
remain qualitatively unchanged. Therefore,
we may consider here only the $l\ge 2$
modes. When we conclude the perturbation analysis we shall give the 
final result for the dipole perturbations too.
                                                              
Let us
take, then, the electromagnetic and gravitational perturbations near the
CH to be \cite{gursel}:
\begin{equation}
Z_{1}^{(+)}=av^{-(2l+2)}+bu^{-(2l+2)}
\end{equation}
\begin{equation}
Z_{2}^{(+)}=cv^{-(2l+2)}+du^{-(2l+2)}.
\end{equation}
We now use the algorithm described in section III to obtain the
electromagnetic field near the CH. 
From $Z_{1,2}^{(+)}$ we can find the functions $H_{1,2}^{(+)}$ by 
Eqs. (C186-C187). 
The function $\Phi$ is given by Eq. (C196). 
Substituting Eq. (C196) in Eqs. (C186-C187), we obtain 
\begin{eqnarray*}
\Phi(r,t)&=&\frac{1}{q_{1}^{2}+|q_{1}q_{2}|}\int_{r}^{r_{-}}
\left\{Z_{1}^{(+)}\left[nr|q_{1}q_{2}|^{\frac{1}{2}}+\sqrt{2n}
Q_{*}q_{1}\right]\right. \nonumber \\
&+&\left. Z_{2}^{(+)}\left[nrq_{1}-Q_{*}\sqrt{2n|q_{1}q_{2}|}\right]\right\}
\frac{e^{-\nu}}{\varpi r}\,dr .
\end{eqnarray*}
As we are interested in the development of the perturbations near the CH, 
we write an approximate expression for $\Phi(r,t)$
to the leading term in $\kappa_{-}r_{*}$.  To do this we expand in 
$r-r_{-}$, and obtain 
\begin{eqnarray}
\frac{e^{-\nu}}{\varpi r}\,dr\approx 
\left[\frac{e^{-\kappa_{-}r_{*}/2}}
{nr_{-}^{2}+3Mr_{-}-2Q_{*}^{2}}+
O(e^{-\frac{3}{2}\kappa_{-}r_{*}})\right]\,dr_{*}. \nonumber
\end{eqnarray}
Substituting in $\Phi$ we get:
\begin{eqnarray}
\Phi(r_{*},t)&\approx & a_{1}\int_{r_{*}}^{\infty} e^{-\frac{1}{2}
\kappa_{-}r_{*}}
\left[(a_{2}a+a_{3}c)v^{-(2l+2)}\right.\nonumber \\ 
&+&\left. (a_{2}b+a_{3}d)u^{-(2l+2)}\right]\,dr_{*},
 \label{phi}
\end{eqnarray}
where
\begin{eqnarray*}
a_{1}&=&\left(q_{1}^{2}+|q_{1}q_{2}|\right)^{-1}
\left(nr_{-}^{2}+3Mr_{-}-2Q_{*}^{2}\right)^{-1} \\
a_{2}&=&nr\sqrt{(-q_{1}q_{2})}+\sqrt{(2n)}Q_{*}q_{1}\\
a_{3}&=&nrq_{1}-\sqrt{(2n)}\sqrt{(-q_{1}q_{2})}Q_{*}.
\end{eqnarray*}
It can be verified (for a formal proof see Ref. \cite{thesis})
that the integral $\int_{x}^{\infty} e^{-gz}z^{-(2l+2)}\,dz\;$ can
be represented by the asymptotic series $$\frac{1}{g}e^{-gx}x^{-(2l+2)}
\sum_{p=0}^{\infty}(-1)^{p}\frac{(p+2l+1)!}{(2l+1)!g^{p}}x^{-p}.$$
Since we are
interested primarily in very large values of $x$, i.e., we are 
interested in the regions of spacetime closest to the CH, 
the zeroth order approximation of the asymptotic series suffices for our 
needs. Therefore, we shall take
$\int_{x}^{\infty}e^{-gz}z^{-(2l+2)}\,dz\approx
\frac{1}{g}e^{-gx}x^{-(2l+2)}$. 
To use this for the evaluation of $\Phi$ we first change the variables in 
Eq. (\ref{phi}), and integrate the two terms separately. Thus, we obtain:
\begin{eqnarray}
\Phi(r_{*},t)\approx 
&-&\frac{2a_{1}}{\kappa_{-}}e^{-\frac{1}{2}\kappa_{-}r_{*}}
\left[(a_{2}a+a_{3}c)v^{-(2l+2)}\right. 
\nonumber \\ &+& \left. (a_{2}b+a_{3}d)u^{-(2l+2)}\right] .
\end{eqnarray}
In order to calculate $B_{03}$, we first need to have $H_{1,r}^{(+)}$. 
[See Eqs. (C190) and (C195).] 
We readily find that  
\begin{eqnarray*}
H_{1,r}^{(+)}&=&\frac{\,dr_{*}}{\,dr}\frac{d}{\,dr_{*}}H_{1}^{(+)}
\nonumber \\ &=&
-(2l+2)\frac{e^{-2\nu}}{q_{1}^{2}+|q_{1}q_{2}|}\left[
\alpha v^{-(2l+3)}+\beta u^{-(2l+3)}\right] ,
\end{eqnarray*}
where $\alpha=q_{1}a-(-q_{1}q_{2})^{1/2}c$ and
$\beta=q_{1}b-(-q_{1}q_{2})^{1/2}d$. We can obtain the
following approximate expression to the dominant term in $e^{-\nu}$ (as
$e^{-\nu}$ diverges on the CH it is clear that the stronger the dependence
of the exponent in $\nu$ the faster the divergence): 
\begin{eqnarray}
B_{03}(r,t)&\approx & -\frac{Q_{*}\mu}{r^{2}}H_{1,r}^{(+)}(r,t)
\nonumber \\ &=&
\frac{(2l+2)Q_{*}\mu}{\Delta\left(q_{1}^{2}+|q_{1}q_{2}|\right)}
\left[\alpha v^{-(2l+3)}+\beta u^{-(2l+3)}\right] . 
 \label{B03(r,t)}
\end {eqnarray}
[In Eq. (\ref{B03(r,t)}) we kept only the leading term in $e^{-\nu}$.]  
Similarly, we also find the dominant term in $e^{-\nu}$
of $B_{23}(r,t)$ near the CH. Namely, in view of Eq. (C190) and 
Eq. (C194)  we get 
\begin{eqnarray*}
B_{23}(r,t)&\approx& -\frac{Q_{*}\mu}{r^{2}}H_{1}^{(+)}(r,t)\\
&=&-\frac{Q_{*}\mu}{r^{2}}\frac{1}{q_{1}^{2}+|q_{1}q_{2}|}
\left[\alpha v^{-(2l+2)}+\beta u^{-(2l+2)}\right] .
\end{eqnarray*}
We thus see that both $B_{03}(r,t)$ and $B_{23}(r,t)$
are the results of linear differential
operators acting on the perturbing fields $Z_{i}^{(+)}$.
As shown in section III [see Eqs. 
(C159)-(C160)] the formalism can be
now used to obtain frequency-dependent tetrad components of the Maxwell 
tensor. As the expressions for the functions $B_{\mu\nu}$ are independent
of the frequency, it is clear that there is a need to adapt the two
representations of the fields (the temporal representation and the
frequency representation). It turns out that
the tetrad components of the Maxwell tensor which lead to divergencies 
are $F_{(0)(3)}(r,\sigma)$ and $F_{(2)(3)}(r,\sigma)$. 

\subsection*{The Components of $F_{(\mu)(\nu)}(r,t,\theta )$ }
Our goal now is to obtain the frequency-independent expression for the
electromagnetic field. As $F_{(0)(3)}(r,\sigma ,\theta )$ 
does not involve the frequency [in view of Eq. (C159)], it is
possible to compute $F_{(0)(3)}(r,t,\theta )$  
directly, without considering the
subtleties of the Fourier transform. 
It can be shown that for the correct performance of the
transformation from the frequency representation to the temporal
representation one should only replace the frequency $\sigma$ of Eq. 
(C160)
with $-id/\,dt$. (For a rigorous proof of this scheme---which is not as 
trivial as it may seem---see Ref. \cite{thesis}.) We thus obtain:
\begin{eqnarray}
F_{(0)(3)}(r,t,\theta)\approx & -&\frac{(l+1)\mu}{q_{1}^{2}+|q_{1}q_{2}|}
\frac{e^{-\nu}}{r}\left[\alpha v^{-(2l+3)}\right. \nonumber \\
&+&\left. \beta u^{-(2l+3)}\right]P_{2,\theta} 
\end{eqnarray}
\begin{eqnarray}
F_{(2)(3)}(r,t,\theta)\approx (l+1)\frac{\mu e^{-\nu}}{r}
\frac{1}{q_{1}^{2}+|q_{1}q_{2}|}\nonumber \\
\times\left[
\alpha v^{-(2l+3)}-\beta u^{-(2l+3)}\right]P_{2,\theta}.
\end{eqnarray}
Here, $P_{l}(\cos\theta )$ denotes the Legendre polynomial of order $l$. 

\section{Derivation of the Maxwell Field Near the 
CH -- Tensorial Components}
This section is built in the following way: We first write down the
tensorial components of the Maxwell tensor in the Schwarzschild 
coordinates, which are known (see, e.g., Ref. \cite{mtw}) to be singular
at the CH. We then transform from the Schwarzschild 
coordinates to Kruskal-Szekeres coordinates, which are regular at
the CH. Finally, we transform from the Kruskal-Szekeres 
coordinates to the rest frame of the infalling object. (The motivation  
behind this last transformation is given below.)
\subsection*{The Schwarzschild Coordinates}
After finding the tetrad components of the electromagnetic field we shall
now find the tensorial components.
From now on we omit the explicit angular dependence of the Legendre
polynomials, to make the expressions simpler and more compact. We may do
this because none of the transformations to be performed below changes the 
angular
coordinates. To allow for this omission we re-define the components of the
Maxwell tensor. Schematically, we separate the variables by  
$F_{(\mu)(\nu)}(r,t,\theta)=F_{(\mu)(\nu)}(r,t)\Theta_{(\mu)(\nu)}(\theta)$.
Hence, from now on $F_{(\mu)(\nu)}(r,t)$ should be understood accordingly.
The tensorial components can be obtained from the tetrad components by the 
transformation  
$F_{\mu\nu}=F_{(\alpha)(\beta)}e^{(\alpha)}_{\mu}e^{(\beta)}_{\nu}$, namely, 
\begin{eqnarray}
F_{03}(r,t)
&=&-re^{\nu}F_{(0)(3)}(r,t)\nonumber \\
&\approx & \frac{(l+1)\mu}{q_{1}^{2}+|q_{1}q_{2}|}
\left[\alpha v^{-(2l+3)}+\beta u^{-(2l+3)}\right]
\end{eqnarray}
\begin{eqnarray}
F_{23}(r,t)
&=&re^{-\nu}F_{(2)(3)}(r,t)\nonumber \\ 
&\approx & \frac{e^{-2\nu}(l+1)\mu}{q_{1}^{2}+
|q_{1}q_{2}|}\left[\alpha v^{-(2l+3)} -\beta u^{-(2l+3)}\right].
\end{eqnarray}
\subsection*{Transforming to Regular Coordinates}
\subsubsection*{Transforming to Kruskal-Szekeres Coordinates}
Now, we wish to transform the Maxwell strength field tensor to a co-moving
reference frame, i.e., a frame in which the infalling observer is at rest.
We do that in order that we could relate the results for different
observers and because of the fact that
to use a (classical or quantum) local theory for the matter-radiation
interaction we need to express the electromagnetic field as measured by 
the physical system in question, and as a function of its local (proper) 
time.
First, we transform from the coordinates 
$(t\; \phi\; r\;\theta )$ 
to the coordinates $(t \; \phi\; r_{*}\; \theta )$. It is clear,
that the only component (out of the two relevant components)
of $F_{\mu \nu}$ which is changed by this
transformation is $F_{23}$. In fact, we get that $$F_{r_{*}\theta}=
e^{2\nu}F_{r\theta}=\frac{(l+1)\mu}{q_{1}^{2}+|q_{1}q_{2}|}
\left[\alpha v^{-(2l+3)}
-\beta u^{-(2l+3)}\right].$$ 
Now, we transform to the coordinates
$(u\; \phi\; v\; \theta )$. (The coordinates $u$ and $v$ are defined in 
section II.) The line element for the unperturbed RN background in
theEddington-Finkelstein coordinates is 
$\,ds^{2}=-\frac{|\Delta |}{r^{2}}\,du\,dv -r^{2}(u,v)\,d\Omega^{2}$. 
We thus find that 
\begin{equation}
F_{u\theta}=F_{r_{*}\theta}-F_{t\theta}=-\frac{2(l+1)\mu}{q_{1}^{2}+
|q_{1}q_{2}|}\beta u^{-(2l+3)}
\end{equation}
\begin{equation}
F_{v\theta}=F_{r_{*}\theta}+F_{t\theta}=\frac{2(l+1)\mu}{q_{1}^{2}+
|q_{1}q_{2}|}\alpha v^{-(2l+3)}.
\end{equation}
Now, we define the Kruskal-Szekeres future directed 
null coordinates $U,V$ by:
\begin{eqnarray}
\ln \left(-\frac{1}{2}\kappa_{-}V\right)=-\frac{1}{2}\kappa_{-}v,
\end{eqnarray}
\begin{eqnarray}
\ln \left(-\frac{1}{2}\kappa_{-}U\right)=-\frac{1}{2}\kappa_{-}u.
\end{eqnarray}
To obtain the form of the metric in these  
coordinates we first need to have an explicit expression for $r_{*}(r)$,
because in transforming from the  
the 
Eddington-Finkelstein coordinates to the
Kruskal-Szekeres coordinates we find for the line element 
\begin{eqnarray}
\,ds^{2}=
-\frac{(r_{+}-r)(r-r_{-})}{r^{2}}e^{\kappa_{-}r_{*}(r)}
\,dU\,dV-r^{2}(U,V)\,d\Omega^{2}. \nonumber
\end{eqnarray} 
As we defined $r_{*}$ in section II only through its differential, 
we realize that when we integrate
to obtain $r_{*}(r)$ we may add an arbitrary integration
constant. This integration constant can be chosen in such a way, that the
line element near the CH assumes a quasi-Minkowskian form\footnotemark
\footnotetext{This is the line element not only exactly on the CH but
also very close to it. (Of course, this line element is regular at the CH.) 
We shall use this quasi-Minkowskian form when we construct
interaction models for the radiation with 
matter (see sections VI and VII).}, 
i.e., $\,ds^{2}=[-1+O(UV)]\,dU\,dV-r^{2}\,d\Omega^{2}$.
Thus, we integrate $\,dr_{*}/\,dr$ and obtain\footnotemark
\footnotetext{Notice the difference between this form of $r_{*}(r)$ and the
forms of Ref. \cite{gursel,chandrasekhar}. Also notice that in
Ref. \cite{chandrasekhar} the definition for the surface gravity
is different from ours.}
\begin{eqnarray}
r_{*}(r)=(r-r_{-})&+&\frac{1}{\kappa_{+}}\ln \frac{r_{+}-r}{r_{+}}
-\frac{1}{\kappa_{-}}\ln \frac{r-r_{-}}{r_{-}}\nonumber \\ 
&+&\frac{1}{\kappa_{-}}
\ln\frac{r_{-}r_{+}^{\kappa_{-}/\kappa_{+}}}{(r_{+}-r_{-})^{1+
\kappa_{-}/\kappa_{+}}}.\nonumber
\end{eqnarray} 
Inserting this in the line element, we find 
\begin{eqnarray}
\,ds^{2}=&-&\left(\frac{r_{-}}{r}\right)^{2}\left(\frac{r_{+}-r}{r_{+}-r_{-}}
\right)^{1+\kappa_{-}/\kappa_{+}}e^{\kappa_{-}(r-r_{-})}\,dU\,dV
\nonumber \\
&-&r^{2}(U,V)\,d\Omega^{2}.
 \label{regular metric}
\end{eqnarray}
Here, $r$ is the implicit function of the coordinates 
$UV=4\exp [-\kappa_{-}r_{*}(r(U,V))]/\kappa_{-}^{2}$.
This metric is of course regular in the domain between the two horizons,
and at the CH, as
should be expected from Ref. \cite{kruskal,graves and brill}.
Due to the regularity of the metric 
(\ref{regular metric}) at the CH, we can 
work
on a sufficiently small neighborhood where spacetime is as close to
Minkowski spacetime as we wish (which is, in fact, trivial, as any
non-singular spacetime possesses this property).
In these coordinates we obtain:
\begin{eqnarray}
F_{V\theta}&=&-\frac{2}{\kappa_{-}V}F_{v\theta}\nonumber\\
&=&-\frac{2(2l+2)}{\kappa_{-}V}
\frac{\mu}{q_{1}^{2}+|q_{1}q_{2}|}\alpha v^{-(2l+3)} 
 \label{one}
\end{eqnarray}
\begin{eqnarray}
F_{U\theta}=-\frac{2}{\kappa_{-}U}F_{u\theta}=\frac{2(2l+2)}{\kappa_{-}U}
\frac{\mu}{q_{1}^{2}+|q_{1}q_{2}|}\beta u^{-(2l+3)}.
 \label{two}
\end{eqnarray}
\subsubsection*{Transformation to the infalling object's rest-frame}
Eqs. (\ref{one},\ref{two}) are not enough for our
needs. The reason for this is two-fold: First, we should like to calculate
the interaction of the electromagnetic field near the CH with the matter
comprising the infalling object. A description in a flat spacetime of
course simplifies the intricate radiative processes considerably. A
flat-spacetime description also allows us to use the standard notions of
electric and magnetic fields. Second, The region where the electromagnetic
field assumes exceptionally high values is a very `narrow' region near the
CH. In this narrow region, because spacetime is approximately
Minkowskian, spacetime curvature is negligible in the background.
Additionally, it will turn out (see sections VI and VII) that
we shall need to describe the interaction of the infalling object with
the radiation field along the entire trajectory, and, in addition, we shall
also wish to compare different observers arriving at totally different
points on the CH. (The reasons for this will become clear in
sections VI and VII.) 
To do that we cannot be satisfied with a completely local
description of the CH,
and therefore we transform now to co-moving coordinates,
which allow us to compare different observers, as for all we set
the proper time equal to zero on arrival to the CH.
Let us now define the coordinates 
$Z,T$ by: $U=Z-T$ and $V=Z+T$. (Note, that
$Z$ is a timelike coordinate and $T$ is a spacelike coordinate.) 
The coordinates $Z,T$ are
Minkowski-like coordinates, yet not appropriate for the description of the
object's rest frame, as generally we should find that the infalling object
is in motion relative to this reference frame.

We now transform to co-moving Minkowski coordinates $\bar{z},\bar{t}$,
adapted to the trajectory 
in question. Namely, we demand that at the
intersection of  the orbit with $V=0$, the newly defined coordinates be
such, that $\dot{\bar{z}}=0$ (and $\dot{\bar{t}}=1$). In addition, we set
$\bar{z}=\bar{t}=0$ at that intersection point.
This transformation is 
thus defined by the coupled equations 
$$ \left \{ \begin{array}{l} 
(1 \; 0 \; 0 \; 0)\equiv \dot{x}^{\alpha '}=\frac{\partial x^{\alpha '}}
{\partial x^{\beta}}\dot{x}^{\beta}\\ \\
\eta^{\alpha '\beta '}\equiv g^{\alpha '\beta '}=\frac{\partial x^{\alpha
'}}{\partial x^{\alpha}}\frac{\partial x^{\beta '}}{\partial x^{\beta}}
g^{\alpha \beta}
\end{array} \right. ,$$
where $\eta^{\alpha \beta}$ is the metric tensor of a Minkowski spacetime
and a dot denotes differentiation with respect to the proper time of the
infalling object.
It is clear, that these transformation equations define the
special-relativistic Lorentz boost
generating the co-moving coordinates uniquely. The solution of these
equations is 
$$\left( \begin{array}{cc}
\frac{\partial \bar{t}}{\partial T} & \frac{\partial \bar{t}}{\partial Z}
\\  &  \\
\frac{\partial \bar{z}}{\partial T} & \frac{\partial \bar{z}}{\partial Z}
\end{array} \right)=\left( \begin{array}{cc}
\dot{T} & -\dot{Z}\\ \\
-\dot{Z} & \dot{T}
\end{array} \right) .$$
We now define the null coordinates $\bar {u}=\bar {z}-\bar {t}$ and 
$\bar {v}=\bar {z}+\bar {t}$. (Note that both $\bar{u}$ and $\bar{v}$
vanish at the intersection of the trajectory 
with $V=0$.) Transforming to these  
coordinates we find
$$F_{V\theta}=\frac{\partial \bar {v}}{\partial V}F_{\bar {v}\theta}+
\frac{\partial \bar {u}}{\partial V}F_{\bar {u}\theta},$$
$$F_{U\theta}=\frac{\partial \bar {v}}{\partial U}F_{\bar {v}\theta}+
\frac{\partial \bar {u}}{\partial U}F_{\bar {u}\theta}.$$
It can be shown that 
$$\frac{\,\partial\bar{u}}{\,\partial V}=0,\;\;
\frac{\,\partial\bar{u}}{\,\partial U}=\dot{V},\;\;
\frac{\,\partial\bar{v}}{\,\partial V}=\dot{U},\;\;
\frac{\,\partial\bar{v}}{\,\partial U}=0.$$
Hence, we get that
$$F_{\bar {v}\theta}=\frac{1}{\dot{U}}F_{V\theta},\;\;\;\;
F_{\bar {u}\theta}=\frac{1}{\dot{V}}F_{U\theta}.$$
It can be readily shown \cite{thesis} that on the CH $\dot{U}(r_{-})=
1/(2\dot{r})e^{-\kappa_{-}u_{0}/2}$ and $\dot{V}(r_{-})=
-2\dot{r}e^{\kappa_{-}u_{0}/2}.$ 
Using this, it can be
shown that on the CH 
$\bar {v}=\dot{U} V$ and $\bar {u}=\dot{V}U-4\dot{r}/\kappa_{-}$. 
After performing all the transformations, we end up with the required
expression for the divergent component of the Maxwell tensor as measured by
an infalling observer in his rest frame. We substitute
$$v=-\frac{2}{\kappa_{-}}\ln \left(-\frac{1}{2}\kappa_{-}\frac{\bar{v}}
{\dot{U}}\right)$$
and
$$u=-\frac{2}{\kappa_{-}}\ln \left[-\frac{1}{2}\kappa_{-}\frac{1}{\dot{V}}
\left(\bar{u}+4\frac{\dot{r}}{\kappa_{-}}\right)\right],$$
and obtain 
\begin{eqnarray}
F_{\bar {v}\theta}&=&-\frac{4(l+1)}{\kappa_{-}\dot{U}V}
\frac{\mu}{q_{1}^{2}+|q_{1}q_{2}|}\alpha v^{-(2l+3)} \nonumber \\
&=&\frac{C'r_{-}}{\kappa_{-}\bar {v}}\left(\ln |\kappa_{-}
\bar {v}|+\frac{1}{2}\kappa_{-}u_{0}+\ln |\dot{r}|\right)^{-(2l+3)},
 \label{Fvtheta}
\end{eqnarray}
\begin{eqnarray} 
F_{\bar {u}\theta} & = &\frac{4(l+1)}{\kappa_{-}\dot{V}U}
\frac{\mu}{q_{1}^{2}+|q_{1}q_{2}|}\beta u^{-(2l+3)} \nonumber \\
& = & -\frac{C''r_{-}}{\kappa_{-}(\bar {u}+\frac{4}{\kappa_{-}}
\dot{r})}\nonumber \\ &\times& \left(
\ln \left|\kappa_{-}\bar {u}+4\dot{r}\right|-\frac{1}{2}
\kappa_{-}u_{0}-\ln |4\dot{r}|\right)^{-(2l+3)},
 \label{Futheta}
\end{eqnarray}
where
\begin{eqnarray*}
C'&=&\frac{4(l+1)\mu}{q_{1}^{2}+
|q_{1}q_{2}|}\alpha\left(\frac{\kappa_{-}}{2}\right)^{2l+3}\frac{1}
{r_{-}} \\
C''&=&\frac{4(l+1)\mu}{q_{1}^{2}+
|q_{1}q_{2}|}\beta\left(\frac{\kappa_{-}}{2}\right)^{2l+3}\frac{1}
{r_{-}}.
\end{eqnarray*}
The form in which we represented 
$F_{{\bar u}\theta}$ might obscure the simplicity of its meaning.
In fact, all we need is the value on the inner horizon, where $\bar{u}=0$.
We readily find that on the CH 
$$F_{{\bar u}\theta}=\frac{l+1}{\dot{r}}\frac{r_{-}\mu}{q_{1}^{2}+
|q_{1}q_{2}|}\beta u_{0}^{-(2l+3)}.$$ From this expression the regularity of
$F_{{\bar u}\theta}$ on the CH is self evident.
\subsubsection*{The final form for the electromagnetic field near the CH
that an infalling observer measures}
We may construe each point of the matter comprising the infalling observer
as being located in the center of its own spatial coordinate system, i.e.,
we may take $\bar{z}=0$ or, equivalently, $\bar{v}=\bar{u}=\bar{t}$.
Identifying the 
coordinate $\bar {t}$ with the observer's proper time $\tau$,
we realize that $\bar{v},\bar{u}$ vanish on the CH.
We get that $F_{\bar{u}\theta}$ remains finite on the CH, while
$F_{\bar{v}\theta}$ diverges.

We can write the required divergent
components of the electric and magnetic fields
near the CH as measured by the infalling observer in an orthonormal
Cartesian tetrad frame in which the observer is at rest.
As the tetrad introduced in section III is orthonormal, it is only 
natural to take here the same tetrad base. Thus, we get:
\begin{eqnarray*}
F_{(\bar{z})(\bar{y})}&=&F_{\bar{z}\theta}e^{\bar{z}}_{(\bar{z})}
e^{\theta}_{(\bar{y})}=F_{\bar{z}\theta}e_{(\bar{y})\mu} g^{\mu\theta}
=\frac{1}{r}F_{\bar{z}\theta},\\
F_{(\bar{t})(\bar{y})}&=&F_{\bar{t}\theta}e^{\bar{t}}_{(\bar{t})}
e^{\theta}_{(\bar{y})}=F_{\bar{t}\theta}e_{(\bar{y})\mu} g^{\mu\theta}
=\frac{1}{r}F_{\bar{t}\theta}.
\end{eqnarray*} 
As the tetrad base is Cartesian and orthonormal,
it is easily shown that both $e^{\bar{z}}_{(\bar{z})}$ and
$e^{\bar{t}}_{(\bar{t})}$ identically equal unity\footnotemark
\footnotetext{A two-dimensional flat Minkowskian spacetime is spanned by
the vectors $\bar{z}$ and $\bar{t}$. As these vectors are orthonormal, the
claim is readily proved.}. We now denote
$E_{\bar{y}}\equiv F_{\bar{t}\theta}$ and
$B_{\bar{x}}\equiv -F_{\bar{z}\theta}$, where 
$F_{\bar{t} \theta}\approx F_{\bar{z}\theta}\approx F_{\bar{v}\theta}$.
Here, $\bar {x}$ and $\bar {y}$ are 
orthonormal tetrad components in the object's 
reference frame, directed in the $\partial / \,\partial\phi$ and the  
$\partial / \,\partial\theta$ directions, respectively. 
Hence, we find that 
$E_{\bar{y}}=-B_{\bar{x}} 
\approx \frac{C'}{\kappa_{-}\tau }
\left( \ln|\kappa_{-}\tau|
+\frac{1}{2}\kappa_{-}u_{0}+\ln |\dot{r}|\right) ^{-(2l+3)}$. 
From these expressions for the divergent components of the electromagnetic
field we may neglect the term dependent on $\dot {r}$, as for typical
observers $\dot {r}$ is neither vanishing nor divergent, and is typically
of order unity. (On the other hand, $\ln |\kappa_{-}\tau |$ diverges, and
$u_{0}$ is typically very large too.) From now on, for the sake of brevity
we shall call these electromagnetic components simply $E$ and $B$,
respectively. Thus, we conclude that
\begin{eqnarray}
E=-B\approx \frac{C'}{\kappa_{-}\tau }
\left( \ln|\kappa_{-}\tau|
+\frac{1}{2}\kappa_{-}u_{0}\right) ^{-(2l+3)}.
 \label{E}
\end{eqnarray}
It is clear from Eq. (\ref{E}) that the electromagnetic 
field---as measured by the infalling observer---diverges. 
This conclusion agrees with the results of Refs.
\cite{penrose,gursel,chandrasekhar,mcnamara,matzner}.

The above analysis was done for $l\ge 2$ polar modes. The analysis can be
repeated for $l=1$ polar modes analogously using the formalism of Ref.
\cite{burko1}. When this is done, it turns out that the electric and
magnetic fields can still be expressed by Eq. (\ref{E}). However, the correct
expression for $C'$ is now 
$C'(l=1)=-8a\left(\kappa_{-}/2\right)^{5}/(q_{1}r_{-})$.

\section{Interaction of the Radiation With Matter: Classical Absorber}
We now consider the interaction of the electromagnetic field 
(\ref{E}) with 
the
matter comprising the infalling object. Here, we consider a classical
object, and in the next section we consider a quantum object. In both the
present and the consecutive sections we assume that the object is much
smaller than the typical 
radius of curvature between the event and the inner horizons, and hence
the effects of curvature are negligible. This arrises from our 
assumption that non-linear effects will not have important effects on the 
radiative interaction of the field with the infalling object. (See the 
discussion in section I.) Consequently, we can 
imagine the object as being at rest in its locally co-moving Minkowski 
frame when an electromagnetic impulse of the shape 
(\ref{E}) comes from null 
infinity and interacts with it. In what follows we shall use the co-moving 
Cartesian coordinates defined in section V, but for simplicity we 
shall omit here the `bars.' Namely, we shall use the coordinates $(\tau 
\;x\;y\;z)$. 

Despite the flatness of spacetime in the 
observer's rest-frame, the interaction of the electromagnetic field with 
the
matter from which the observer is made is extremely complicated. Therefore,
we take a very simplified 
(toy-) model for the matter-field interaction, 
which
still embodies the most essential properties of more realistic 
interactions.
We take the matter to be composed of classical ``atoms." Each ``atom" is
composed of two point-like oppositely charged
particles, with charges $\pm e$ and masses $\mu_{\pm}$,
correspondingly. We denote the reduced
mass by $\mu$. (Do not confuse $\mu$
here, which is the reduced mass, with $\mu$ in the previous two sections.) 
The system interacts with an external force, which in our case is the
Lorentz force induced by the blue-shifted incoming electromagnetic field.
This external force acts to change the separation distance between the two 
particles of the ``atom." We presume small deviations from equilibrium (to 
be
justified {\it a posteriori}), and hence take a linear restoring force,
i.e., $F=-\mu\omega^{2}X$, where $X$ is the deviation (of the particles'
separation) from equilibrium, and $\omega$ is the resonance frequency.
The dipole is chosen to be aligned in the
$\partial /\,\partial\theta$ direction (to allow for a maximum
interaction with the field). We take the initial conditions to be $X=0$
and $\dot{X}=0$. The excitation of the system by the
field may be characterized by $X$, $\dot{X}$, and also by
the total absorbed mechanical energy
$\cal{E}_{\rm c}$. We shall show that all three variables are finite,
small, and for typical parameters even negligible.

The equation of motion is
\begin{eqnarray}
\mu\ddot{X}+\mu\omega^{2}X=eE(\tau),
\end{eqnarray}
where $E(\tau)$ is the divergent component of the electric field 
(\ref{E}).
(The contribution of the magnetic field is neglected, as the ratio of the
electric term to the magnetic term in the expression for the Lorentz
force is proportional to the system's internal velocity $\dot{X}$, which
is taken to be small---a presumption which is justified 
{\it a posteriori}.)

The solution of this equation (with our initial conditions) is
\begin{eqnarray}
X(\tau )&=&
-\frac{1}{2i\omega}\frac{e}{\mu}e^{-i\omega\tau}\int_{-T}^{\tau }
E(\tau ')e^{i\omega\tau '}\,d\tau '+{\rm c.c.} 
 \label{X}
\\
\dot{X}(\tau)&=&\frac{1}{2}\frac{e}{\mu}e^{-i\omega\tau}\int_{-T}^{\tau }
E(\tau ')e^{i\omega\tau '}\,d\tau '+{\rm c.c.},
 \label{V}
\end{eqnarray}
where $T$ is the time of infall from the event horizon to the CH.
Calculating the sum of the
kinetic and potential energies, we find for the total
absorbed mechanical energy
\begin{eqnarray}
{\cal E}_{\rm c}(\tau)&=&\frac{1}{2}\mu\omega^{2}X^{2}+\frac{1}{2}\mu
\dot{X}^{2}\nonumber \\
&=&\frac{1}{2}\mu\left(\frac{e}{\mu}\right)^{2}\left|\int_{-T}^{\tau}E(\tau 
')
e^{i\omega\tau '}\,d\tau '\right|^{2}.
 \label{Energy}
\end{eqnarray}
In all of the three expressions (\ref{X}),(\ref{V}), and (\ref{Energy}) 
we need to calculate the integral 
\begin{eqnarray}
I=\int_{-T}^{\tau}E(\tau ')e^{i\omega\tau '}\,d\tau '.
\end{eqnarray} 
The evaluation of this integral for typical parameters is done explicitly 
in Appendix A. The evaluation yields [Eq. (A27)],
to the leading orders in $(\kappa_{-}u_{0})^{-1}$
\begin{eqnarray}
I(\tau =0)\approx&-&\frac{C'}{(2l+2)\kappa_{-}}\left(\frac{1}{2}\kappa_{-
}u_{0}
\right)^{-(2l+2)}\nonumber \\
&+&\frac{C'}{\kappa_{-}}\frac{1}{2}\pi i
\left(\frac{1}{2}\kappa_{-}u_{0}\right)^{-(2l+3)},
\end{eqnarray}
Substituting in the explicit expressions for $X$, $\dot{X}$,
and $\cal{E}_{\rm c}$ we obtain, to the leading order in $(\kappa_{-}
u_{0})^{-1}$
\begin{eqnarray}
{\cal E}_{\rm c}(0)&\approx&\frac{1}{2(2l+2)^{2}}
\frac{C'^{2}}{\kappa_{-}^{2}}\mu
\left(\frac{e}{\mu}\right)^{2} 
\left(\frac{1}{2}\kappa_{-}u_{0}\right)^{-2(2l+2)},
 \label{E0}\\
\dot{X}(0)&\approx&\frac{1}{2l+2}\frac{C'}{\kappa_{-}}\left(\frac{e}
{\mu}
\right)\left( \frac{1}{2}\kappa_{-}u_{0}\right)^{-(2l+2)}, 
 \label{V0}\\
X(0)&\approx&
-\frac{1}{2}\pi \frac{C'}{\kappa_{-}\omega}\left(\frac{e}{\mu}\right)
\left( \frac{1}{2}\kappa_{-}u_{0}\right)^{-(2l+3)}. 
 \label{X0}
\end{eqnarray}
We find that the strength of the excitation depends on $u_{0}$.
However, $u_{0}$ cannot be directly evaluated by an outside observer, who
wishes to predict the excitation strength should he jump into the BH. 
Thus, it would be worthwhile
to express the excitation strength in terms of the
{\it external} parameters of the problem.
This can be done once we realize that 
$\,du_{0}/\,dv_{0}=-1$. 
(The proof is given in Ref. \cite{thesis}.) 
We find that the infalling observer can
increase $|u_{0}|$ by simply waiting outside the BH before
jumping in and thus increasing the value of $v_{0}$. For a sufficiently
large $|u_{0}|$ we get that ${\cal E}_{\rm c}(0)$, $X(0)$ and $\dot{X}(0)$ 
are
finite and arbitrarily small.

We conclude that despite the divergence of the electromagnetic field, the
excitation of the classical system (and the energy absorbed) is finite, and
becomes arbitrarily small, for late-time observers (large $v_{0}$).
Therefore, the behavior of the charged classical system
obtained by the above
analysis is regular, and despite the divergence on the CH of the external
force acting on the system, the energy absorbed by it is finite and 
negligible for a sufficiently large $v_{0}$.

\section{Interaction of the Radiation With Matter: Quantum Absorber}

As a quantum analogue to the preceding classical system we take
(again) a very simplified model, which, we believe, captures the
essential properties for the interaction we study and
neglects all irrelevant details.
Let us deal then with a non-degenerate quantum system obeying
the Schr\"{o}dinger equation. We shall evaluate the excitation of the
system by considering the transition of the system from its ground state to
an excited state. When there are many (or even an infinite number of)
excited states, our considerations can be generalized for the analysis
of the excitation. In the following we shall discuss, then, only the
excitation between two quantum states.
The unperturbed eigenstates of the system are the ground state  
$|\psi_{i}\rangle e^{-i\omega_{i}t}$ and the excited state
$|\psi_{f}\rangle e^{-i\omega_{f}t}$. 
(That is,
the eigenstates of the quantum system when there is no electromagnetic
field due to the blue sheet. This could be, for instance, the eigenstates
of the system in its original orbit around the BH before the jump
in, or its eigenstates when the system crosses the event horizon.) 
Namely, the obvious evolution in time is given by the standard oscillatory
dependence, and then $|\psi_{n}\rangle$ is independent of the time, where
$n=i,f$. The perturbed wave function would be given then by
\begin{eqnarray}
|\psi\rangle =\sum_{n}a_{n}|\psi_{n}\rangle e^{-i\omega_{n}t},
\end{eqnarray}
where $a_{n}$ are the expansion coefficients.
The system is taken initially at its ground state $|\psi_{i}\rangle$.
Therefore, we
take $a_{i}^{\rm initial}=1$ and $a_{f}^{\rm initial}=0$. Assuming small
transitions (an assumption which will be justified {\it a posteriori}),
we take $a_{i}^{\rm final}\approx 1$ too. The system is perturbed by the
pulse of the electromagnetic field. In the Coulomb gauge, which is 
consistent
with our treatment, the interaction Hamiltonian is
${\cal H}=(e/m){\bf A\cdot p}$,
where we keep only linear terms in the perturbation, which
is assumed to be small. Here, ${\bf A}$ is the electromagnetic
vector-potential written in the Coulomb gauge, and ${\bf p}$ is the 
3-momentum.
(The explicit form of ${\bf A}$ is not important for our needs here,
although it can be found from ${\bf E}=-\,\partial 
{\bf A}/\,\partial \tau-{\bf \nabla}\phi$ and 
${\bf B}={\bf \nabla}\times {\bf A},$
where $\phi$ is the temporal component of the four-potential.) It can be
easily verified that any vector potential of the form
${\bf A}=A(\bar{v}){\bf e}_{y},$
where ${\bf e}_{y}$ is a unit vector in the $y$ direction,
is consistent with these expressions and with Eq. (\ref{E}). It can be
further shown that this form for the vector potential is consistent also
with the Coulomb gauge condition, namely, with ${\bf \nabla\cdot A}=0$.
In what follows, we take the temporal component of the four-potential to
vanish. This can be done consistently with the Coulomb gauge. The vector
potential can thus
be written explicitly as ${\bf A}=-\int {\bf E}(\tau ')\,d\tau '.$ 

We are interested in the effects of the blue sheet, namely, with the
effects of the pulse of the  electromagnetic field on our system.
Therefore, we shall assume, for the sake of brevity and simplicity, that 
the
electromagnetic pulse ends at the CH, or, in other words, that
for {\it each} point of the system the electric and the magnetic fields
vanish for positive proper time. More accurately, if the system's spatial
extension in the $z$ direction is $\,\delta z$ from the center of the 
system, we shall examine the excitation of the system at proper time
(of the system's central point) $\tau >\,\delta z>0$. More conveniently, we
shall look at the system's state at $\bar{v}>0$, when there is no
electromagnetic field associated with the blue sheet. 
[Of course, as physics {\it beyond} the CH is as yet unknown,
we do not suggest here that there are no perturbing fields on the other side 
of the CH. Our point here, is that for the sake of the 
calculation of the blue-sheet effects, the specific form of the fields 
beyond the CH is irrelevant. Moreover, our choice here (to set the 
electromagnetic field equal to zero for positive $\bar{v}$) is no worse  
than
any other choice (in view of the present knowledge of the physics beyond the
CH).] 
We shall calculate the energy absorbed by the system, as a measure for the
strength of its excitation due to the divergent electromagnetic field.
By first order time-dependent perturbation theory $a_{f}^{\rm final}$ is 
given
by
\begin{eqnarray}
a_{f}^{\rm final}=-\frac{i}{\hbar}\frac{e}{m} \int_{-\infty}^{\tau}\langle
\psi_{f}\left| {\bf A\cdot p}\right|\psi_{i}\rangle e^{i\Omega_{fi}\tau '}
\,d\tau ',
\end{eqnarray}
where $\Omega_{fi}=\omega_{f}-\omega_{i}$. 
Integrating by parts, we find that
\begin{eqnarray}
a_{f}^{\rm final} &=&-\frac{e}{m}\frac{1}{\hbar\Omega_{fi}}e^{i\Omega_{fi}
\tau} \langle\psi_{f}\left|{\bf A\cdot p}\right|\psi_{i}\rangle (\tau )
\nonumber \\
&+&\frac{e}{m}\frac{1}{\hbar\Omega_{fi}}\int_{-\infty}^{\tau}
\left\langle\psi_{f}\left| \partial {\bf A}/\partial t{\bf \cdot p}
\right|\psi_{i}\right\rangle e^{i\Omega_{fi}t}\,dt\nonumber \\
&=&-\frac{e}{m}\frac{1}{\hbar\Omega_{fi}}e^{i\Omega_{fi}\tau}\langle\psi_{f}|
{\bf A\cdot p}|\psi_{i}\rangle (\tau )\nonumber \\
&-&\frac{e}{m}\frac{1}{\hbar\Omega_{fi}}\int_{-\infty}^{\tau}\langle\psi_{f}|
{\bf E\cdot p}|\psi_{i}\rangle e^{i\Omega_{fi}t}\,dt \nonumber \\
&\equiv & a_{fi}^{(1)}+a_{fi}^{(2)}.
 \label{Q3}
\end{eqnarray}
The first term does not contribute to the physical excitation. The problem
with $a_{fi}^{(1)}$ is that apparently one could (incorrectly) infer that
even {\it after} the perturbation vanishes, $a_{f}^{\rm final}$ continues
to evolve. To be more specific, it looks as though the system is being
perturbed (and excited) even when there is no perturbation at all.  
This 
is, of course, an erroneous result. We shall resolve this ``paradox'' in
Appendix B, and conclude that $a_{fi}^{(1)}$ describes a pure gauge
distortion of the wave-function, which
can be set equal to zero by a proper gauge transformation.
That is, when we adjust the gauge such that ${\bf A}$ 
vanishes 
for $\tau >0$, 
the above problematic term simply disappears. Therefore,
we are thus left
only with the gauge-independent energy absorption
${\cal E}_{\rm q}(\tau)$ which is given by
\begin{eqnarray}
{\cal E}_{\rm q}(\tau)
&=& \left| a_{fi}^{(2)} \right|^{2}\hbar\Omega_{fi}\nonumber
\\
&=&\frac{1}{\hbar\Omega_{fi}}\frac{e^{2}}{m^{2}}\left|
\int_{-\infty}^{\tau}\langle\psi_{f} |{\bf E(\tau ,{\bf r})
\cdot p}|\psi_{i}\rangle
e^{i\Omega_{fi}t}\,dt\right| ^{2}\nonumber \\
&=& \frac{1}{\hbar\Omega_{fi}}\frac{e^{2}}{m^{2}}\left|
\left\langle\psi_{f}\left|\left(\int_{-\infty}^{\tau}e^{i\Omega_{fi}t}{\bf E}
\,dt\right){\bf \cdot p}\right|\psi_{i}\right\rangle\right|^{2}.
 \label{QE}
\end{eqnarray}
The physical energy-absorption we are interested in is the energy 
absorption just after the system has fully crossed the CH, in
the meaning described previously. As we are interested here in the
effects of the blue sheet, let us suppose then that after the CH the
electric field vanishes, and therefore the quantum
system does not undergo any further excitation. The integrand in Eq. 
(\ref{QE})
vanishes for $\bar{v}>0$, and therefore we may change the variables in the
integration, and get
\begin{eqnarray}
{\cal E}_{\rm q}(\bar {v}=0)=
\frac{1}{\hbar\Omega_{fi}}\frac{e^{2}}{m^{2}}
\left|\left\langle\psi_{f}\left|e^{-i\Omega_{fi}\bar{z}}
\right.\right.\right.\nonumber\\
\left.\left.\left. \times  \left(\int_{-\infty}
^{0}e^{i\Omega_{fi}\bar{v}}E(\bar{v})\,d\bar{v}\right)p_{y}
\right|\psi_{i}\right\rangle\right|^{2}\nonumber \\
=\frac{1}{\hbar\Omega_{fi}}\frac{e^{2}}{m^{2}}\left|\left\langle
\psi_{f}\left| e^{-i\Omega_{fi}\bar {z}}p_{y}\right|\psi_{i}\right
\rangle\right| ^{2}\nonumber\\ 
\times  \left|\int_{-\infty}^{0}e^{i\Omega_{fi}\bar {v}}
E(\bar {v})\,d\bar{v}\right| ^{2},
 \label{QE1}
\end{eqnarray}
where $p_{y}={\bf p\cdot e}_{y}$.
The integral in Eq. (\ref{QE1}) is the same integral
as in the expression for the
total mechanical energy of the classical oscillator at the CH, 
namely, the integral in Eq. (A1).
Therefore, using Eq. (A27)
we get that to the leading
order in $(\kappa_{-}u_{0})^{-1}$
\begin{eqnarray}
{\cal E}_{\rm q}(0)&=&\frac{1}{(2l+2)^{2}}
\frac{1}{\hbar\Omega_{fi}}\left(\frac{e}{m}\right)^{2}
\frac{C'^{2}}{\kappa_{-}^{2}}\left|\langle\psi_{f}|
e^{-i\Omega_{fi}\bar{z}}p_{y}|
\psi_{i}\rangle\right|^{2}\nonumber \\ 
&\times & \left(\frac{1}{2}\kappa_{-}u_{0}\right)
^{-2(2l+2)}.
 \label{Eq0}
\end{eqnarray}
The matrix element in Eq. (\ref{Eq0}) is regular and does not diverge on
the CH any more than anywhere else. We also see the close
correspondence between the
classical and the quantum systems. As the matrix element in Eq. (\ref{Eq0}) 
has dimensions of
momentum squared, we see that the energy gap between the states
times the probability amplitude for the quantum system indeed has the
dimensions of mass. In the classical system the absorbed energy is equal to
the product of half the reduced mass of the system and the square of the
internal velocity of the system.  We see, that the two expressions for the 
absorbed energy in the two systems indeed correspond. The advantage of the
classical treatment is that it does not involve perturbation theory. It
suffers, though, from the fact that it is a classical model, while actual
physical matter is intrinsically quantum. Thus,
the two models contribute to the understanding of each other, and augment
our understanding of the interaction of the blue sheet inside a RN BH 
with
infalling objects. 

\section{Discussion}
In this Paper we investigated the following question: Are physical 
objects 
necessarily burnt up by the blue sheet inside a BH? This is a key
question for a more complete understanding of the internal structure of
BHs and for an understanding of the possibility to fall into a 
BH and re-emerge in another universe. To address this question, we
analyzed the interaction of the blue sheet with an infalling physical
object. We first derived an explicit expression for the electromagnetic
field as measured in the rest frame of the infalling observer, and then we
modeled the interaction of the blue sheet with the observer 
in two ways: a classical
model and a quantum model. In both models we calculated a measure for the
strength of the interaction of the blue sheet with the infalling object. 

We have shown, that the divergence of the energy density of the radiation
(or, even, the divergence of the integral of this energy density 
over proper time) at the CH does not mean that any
physical object falling into a BH will necessarily be completely burnt up
by the radiation. Even though we toy-modeled the matter comprising the
infalling object in a very simplified way, we believe that our models
capture the essence of the interaction of radiation with matter. Therefore,
we conclude that the classical Maxwell radiation created during the 
collapse of the star will not necessarily
destroy objects at the CH. We find, that
the interaction of the blue sheet with physical objects is finite.
Moreover, for typical parameters of astrophysical supermassive BHs
this interaction is even arbitrarily small. In fact, one can diminish the
extent 
of the interaction by simply waiting outside the BH before jumping into
it. This means that if a spaceship is in orbit around the BH, and
an astronaut wishes to fall into it (hoping to re-emerge in another
universe), he should just wait in the spaceship, and postpone the beginning
of his unusual odyssey. According to our analysis, the longer he waits, the
smaller the interaction, and the safer the voyage.

We should
remember though, that throughout this Paper we have ignored other 
inherent radiation sources,
such as the cosmic background radiation. In addition, a more realistic
treatment of the interaction of the blue sheet with infalling objects
will have to consider {\sc Qed} effects, which we have ignored here
completely. Such {\sc Qed} effects, and in particular effects caused by
pair-production, are expected to change the general picture portrayed by 
our analysis considerably. These quantum effects
might be crucial for a more complete understanding of the
blue sheet and its interaction with infalling objects. We showed  
elsewhere, that these {\sc Qed} effects could be fatal for a
human-being observer (due to his high vulnerability to $\gamma$ rays),
but typical physical objects of similar or smaller size
might survive it. These results do not provide support to the idea that no 
continuation of the geometry beyond the CH is physically reasonable 
\cite{burko3}. 

\section*{Acknowledgements}
I am indebted to Amos Ori for considerable help and much good advice. 

\section*{Appendix A: Evaluation of the Integral}

In order to evaluate the integral
$$
I=\int_{-\infty}^{\tau}e^{i\omega\tau '}E(\tau ')\,d\tau '
\eqno{\rm (A1)}
$$
let us divide the
integration region into three qualitatively different regions, denoted by 
$a
,b$ and $c$, respectively. In region $a$ we assume that
$-\infty<\tau\ll-\omega^{-1}$. Since the variation of $E$ with $\tau$ is
dominated by $1/\tau$, in region $a$
the electric field changes very slowly compared with the
exponential in the integrand. Hence, we shall approximate the evaluation of
the integral by taking the electric field outside the integration.
We thus obtain that $I_{a}\approx e^{i\omega\tau}E(\tau)/(i\omega)$.
We now evaluate the error associated with
this approximation: Integrating by parts (in region $a$), we obtain
$$
I_{a}=\frac{1}{i\omega}\left[e^{i\omega \tau}E(\tau )-\int_{-\infty}^{\tau}
e^{i\omega \tau '}\frac{\,dE}{\,d\tau '}\,d\tau '\right], \eqno{\rm (A2)}
$$
or 
$$
I_{a}\approx \frac{1}{i\omega}\left[ e^{i\omega \tau}E(\tau )-
\frac{1}{i\omega}\frac{\,dE}{\,d\tau}\int_{-\infty}^{\tau}e^{i\omega \tau 
'}
\,d\tau ' \right] . \eqno{\rm (A3)}
$$
Hence, the error in the evaluation of $I_{a}$ is:
$$|\Delta I_{a}|\approx \left|\frac{1}{i\omega}e^{i\omega \tau}\left[
E(\tau )-\frac{1}{i\omega}\frac{\,dE}{\,d\tau}-E(\tau )\right]\right|=
\frac{1}{\omega^{2}}\left|\frac{\,dE}{\,d\tau}\right| .\eqno{\rm (A4)}$$
As we can take (for region $a$) $E(\tau )\approx C'/(\kappa_{-}
\tau)$, we get that
$$|\Delta I_{a}|\approx \frac{C'}{\kappa_{-}}
\frac{1}{(\omega\tau )^{2}}. \eqno{\rm (A5)}$$
The relative error is, thus,
$$\left|\frac{\Delta I_{a}}{I_{a}}\right|\approx \left|\frac{\frac{1}
{(\omega\tau )^{2}}}{\frac{1}{\omega\tau}}\right|=
\left|\frac{1}{\omega\tau}\right|.
\eqno{\rm (A6)}$$
We choose now for the proper time $\tau$ at the
boundary between regions $a$ and $b$ the following value:
$$\tau_{a}\approx (\kappa_{-}\omega)^{-1/2}, \eqno{\rm (A7)}$$
and therefore we obtain 
$$\left|\frac{\Delta I_{a}}{I_{a}}\right|\approx
\sqrt{\frac{\kappa_{-}}{\omega}}\ll 1. \eqno{\rm (A8)}$$
[For typical parameters (see below) $\; \kappa_{-}/\omega$ 
is of order $10^{-29}$.] 
Hence, we find that the approximation we take for region $a$ is valid.

Region $c$ is defined by $-\omega^{-1}\ll
\tau <0$, i.e., by the requirement that the electric field varies very fast
compared with the exponential term, and therefore we can take the latter
outside the integration. Before we perform this explicitly let us deal with
region $b$, where most of the problems lie.

Region $b$ is in between regions $a$ and $c$. The main difficulty in the
evaluation of the contribution of region $b$ to the integral (A1) comes
from the neighborhood of $\tau\approx -\omega^{-1}$: In that region neither
of the two approximations (the one for region $a$ and the one for region
$c$) is valid. It turns out, however, that in that region there is a
different approximation we can use; the logarithm-dependent term in the
electric field (\ref{E}) does not change much in
comparison with the $1/\tau$ term in region $b$. It turns out, that if we
assume that $\omega^{-1}\ll M\ll -u_{0}$ (which is a very plausible
assumption for physical BHs), 
region $b$ overlaps with both regions $a$ and
$c$. Thus, using the three different approximations (for regions $a,b$, and
$c$), we can calculate the integral (A1) for the entire interval
$-\infty < \tau < 0$. In fact, the matching between regions $a$ and $b$ is
done automatically because of the combination of two facts:
first, in region $a$ the
integral follows the electric field adiabatically, and does not have a
`memory' of its values in former times; second, region $b$ overlaps with
region $a$. Therefore, the integral (A1) assumes the form
$$I=\frac{E_{0}}{\kappa_{-}}
\int_{-\infty}^{\tau}\frac{1}{\tau '}e^{i\omega\tau '}\,d\tau '
\equiv \frac{E_{0}}{\kappa_{-}}I_{b} , \eqno{\rm (A9)}$$
where $E_{0}=C'\left(\ln |\kappa_{-}\omega^{-1}|+\frac{1}{2}\kappa_{-}
u_{0}\right)^{-(2l+3)}$. We justify the protraction of the lower limit of
the integration interval from $-T$ to $-\infty$ by the vanishing value of
this expression in the limit of $\tau\longrightarrow -\infty$.
Region $b$ extends up to $\tau =\tau_{b}$. We take $\tau_{b}$ such that
$\vartheta=\omega |\tau_{b}|$, where $\vartheta\ll 1$ is a dimensionless
constant, to be evaluated. (In Ref. \cite{thesis}
we calculate an optimal value of the boundary between
regions $b$ and $c$, i.e., we calculate $\vartheta$. 
However, this optimal value for $\vartheta$ is of no crucial importance. 
Yet, it turns out that this optimal value is 
$\vartheta\approx-(2l+3)(\kappa_{-}u_{0}/2)^{-1}\ln |
(2l+3)(\kappa_{-}u_{0}/2)^{-1}|$.) 
This means
that both approximations $a$ and $b$ are valid near $\tau =\tau_{b}$. We 
write the integral in Eq. (A9) as
$$\int_{-\infty}^{\tau_{b}}\frac{1}{\tau}e^{i\omega\tau}\,d\tau=
\int_{-\infty}^{\vartheta}\frac{1}{\omega\tau}e^{i\omega\tau} 
\,d(\omega\tau). \eqno{\rm (A10)}$$ 
[In the approximation for region $b$ we suppose that 
$\omega ^{-1}\ll M\ll -u_{0}$: 
we take the BH mass to be $10^{9}M_{\odot}$, where $M_{\odot}$ denotes the 
solar mass. (Even 
though a typical mass for a supermassive galactic BH may be 
``only'' about $10^{7}$--$10^{8}M_{\odot}$, this does not affect our 
analysis.) This mass is equivalent to 
a time period of $5\cdot 10^{3}$ seconds (as $1M_{\odot}$ is equivalent 
to 5 $\mu {\rm sec}$). We also take the time of the jump into the BH to 
be of the order of magnitude of a typical cosmological time scale, e.g., we 
take $u_{0}=-10^{9}$ years. This means that $\kappa_{-}u_{0}$ is of the 
order of $-6\cdot 10^{12}$. If we take $\omega$ to be of the order of 
$10^{16}\; {\rm sec}^{-1}$, which is a typical angular frequency for 
atomic processes, we get that $\ln | \kappa_{-}\omega^{-1}| \approx -40$.  
The typical infall time for a BH $T\approx M$, and 
therefore $\ln |\kappa_{-}T|\approx 1$. This means that the variation in  
the logarithmic dependent term throughout the protracted 
region $b$ is negligible in comparison with the magnitude 
of $\kappa_{-}u_{0}$, which justifies the approximation we made for region 
$b$.] 
Hence, we get that in region $b$ the integral becomes:
$$
I_{b}=\int_{-\infty}^{\omega\tau_{b}}\frac{\cos\omega\tau}{\omega\tau}
\,d(\omega\tau)+i\int_{-\infty}^{\omega\tau_{b}}\frac{\sin\omega\tau}
{\omega\tau}\,d(\omega\tau)  , \eqno{\rm (A11)}$$
or, more conveniently,
$$I_{b}\equiv \left(I_{1}+iI_{2}\right). \eqno{\rm (A12)}
$$
We now calculate each term separately:
$$I_{2}=
\int_{-\infty}^{0}\frac{\sin y'}{y'}\,dy'+\int_{0}^{y}\frac{\sin y'}
{y'}\,dy' =\frac{1}{2}\pi+{\rm Si}(y), \eqno{\rm (A13)}$$
where ${\rm Si}(\tau)$ is the sine integral defined by 
${\rm Si}(z)=\int_{0}^{z}\frac{\sin t}{t}\,dt$ \cite{abram} and 
$y=\omega\tau_{b} .$ As in the overlap region
between regions $b$ and $c$ we have $-\omega\tau\ll 1$, the sine integral
is much smaller than unity, and we thus obtain that
$$I_{2}\approx \frac{1}{2}\pi. \eqno{\rm (A14)}$$
Turn now to $I_{1}$:
\begin{eqnarray*}
I_{1}&=&\int_{-\infty}^{y}\frac{\cos y'}{y'}\,dy'= 
\int_{-\infty}^{-1}\frac{\cos y'}{y'}\,dy'+
\int_{-1}^{y}\frac{\cos y'}{y'}\,dy',
\end{eqnarray*}
or, 
$$
I_{1} \equiv  I_{11}+I_{12}.
\eqno{\rm (A15)}$$
It is straightforward to show that the integral $I_{11}$ is bounded and of
order unity. An explicit calculation yields\footnotemark \footnotetext{
In fact, $I_{11}={\rm Ci}(1)$, where ${\rm Ci}(z)$ is the cosine integral
defined by ${\rm Ci}(z)=\gamma +\ln z +\int_{0}^{z}[(-1+\cos t)/t]
\,dt$, where $\gamma$ is Euler's constant \cite{abram}.}:
$$I_{11}=\int_{-\infty}^{-1}\frac{\cos y'}{y'}\,dy'\approx
0.3374. \eqno{\rm (A16)}$$
We now write the integral $I_{12}$ as:
$$I_{12}=
\int_{-1}^{y}\frac{-1+\cos y'}{y'}\,dy'+
\int_{-1}^{y}\frac{1}{y'}\,dy',$$
or, 
$$I_{12}
=\int_{-1}^{y}\frac{-1+\cos y'}{y'}\,dy'
+\ln |y|. \eqno{\rm (A17)}$$
The integral on the right-hand side of (A17) can be written as:
$$\int_{-1}^{y}\frac{-1+\cos y'}{y'}\,dy' = A+O(y^{2}), \eqno{\rm (A18)}$$
where $A\equiv \int_{-1}^{0}(-1+\cos y')/y'\,dy'$. Again, it is
straightforward to show that the integral A is bounded and of order unity,
and numerical calculation yields\footnotemark \footnotetext{
In fact, $A=\sum_{p=1}^{\infty}(-1)^{p+1}\frac{1}{2p(2p)!}$.}
$A\approx 0.2398$.
Therefore, we get that 
$$I_{b}\approx K+\ln|y|+O(y^{2})+i\left[\frac{\pi}{2}+O(y)
\right], \eqno{\rm (A19)}$$
where $K\equiv {\rm Ci}(1)+A\approx 0.5772,$ and where we kept only the
leading terms. We see from Eq. (A19) that $I_{b}$ contributes to both the
real and the imaginary parts of $I$. However, we now explicitly see the
reason for which region $b$ can not be protracted all the way to $y=0$: the
logarithmic term in ${\rm Re}(I_{b})$ diverges as $y\longrightarrow 0$. In
Ref. \cite{thesis} 
we calculate $\omega\tau_{b}$. It is shown there, that when
the value of $y$ at the boundary between regions $b$ and $c$ is taken for
the evaluation of the logarithmic term in Eq. (A19), the contribution of
$I_{b}$ to the real part of $I$ is negligible in comparison with the
contribution of $I_{c}$ to the real part of $I$.
Therefore, we obtain that on the boundary of regions $b$ and $c$, $I$ is
proportional to $(\kappa_{-}u_{0})^{-(2l+3)}$ due to the proportionality to
$E_{0}$. Let us now evaluate the error in our approximation, and thus show
that the approximation is valid. The exact value of the integral (A1) in
region $b$ is
\begin{eqnarray}
I_{b}^{\rm exact}&=&\int_{\tau_{a}}^{\tau_{b}}\frac{1}{\tau}
\left(\ln |\kappa_{-}\tau |+\frac{1}{2}\kappa_{-}u_{0}\right)
^{-(2l+3)}e^{i\omega\tau}\,d\tau\nonumber \\
&=&\left(\ln |\kappa_{-}\tau_{b}|+\frac{1}{2}\kappa_{-}u_{0}\right)^{-(2l+
3)} 
\int_{\tau_{a}}^{\tau_{b}}\frac{1}{\tau}e^{i\omega\tau}\,d\tau\nonumber \\
&-&(2l+3)\int_{\tau_{a}}^{\tau_{b}}\left(\int^{\tau}
\frac{1}{\tau}e^{i\omega\tau}\,d\tau \right)\nonumber \\
&\times & 
\left(\ln |\kappa_{-}\tau |+\frac{1}{2}\kappa_{-}u_{0}\right)^{-(2l+4)}
\,d\tau.
\nonumber
\end{eqnarray}
Our approximation is
\begin{eqnarray}
I_{b}^{\rm approx.}&=&\left(\ln |\kappa_{-}\tau_{b}|+\frac{1}{2}
\kappa_{-}u_{0}\right)^{-(2l+3)}
\int_{\tau_{a}}^{\tau_{b}}\frac{1}{\tau}e^{i\omega\tau}\,d\tau . \nonumber
\end{eqnarray}
Therefore, the relative error is
\begin{eqnarray}
\left| \frac{I_{b}^{\rm approx.}-I_{b}^{\rm exact}}{I_{b}^{\rm approx.}}
\right| \approx
(2l+3)\nonumber \\ 
\times  \left|\frac{1}{\ln |\kappa_{-}\tau_{b}|+\frac{1}{2}\kappa_{-}
u_{0}}\frac{\int_{\tau_{a}}^{\tau_{b}}\left(
\int^{\tau}\frac{1}{\tau '}e^{i\omega\tau '}\,d\tau ' \right)\frac{1}{\tau}
\,d\tau}{\int_{\tau_{a}}^{\tau_{b}}\frac{1}{\tau}e^{i\omega\tau}\,d\tau}
\right| \nonumber \\ 
\approx (2l+3)\left|\frac{\ln |\omega\tau_{b}|}
{\ln |\kappa_{-}\tau_{b}|+\frac{1}{2}\kappa_{-}u_{0}}\right|.\nonumber
\end{eqnarray}
Taking now the optimal value for the boundary between regions $b$ and $c$
(see Ref. \cite{thesis}), we obtain
$$\left|\frac{\Delta I_{b}}{I_{b}}\right|\approx
(2l+3)\left|\frac{\ln \left|\frac{1}{2}\kappa_{-}u_{0}\right|}
{\frac{1}{2}\kappa_{-}u_{0}}\right|\ll 1. \eqno{\rm (A20)}$$
Thus, our approximation for region $b$ is valid.

Region $c$ is defined by the requirement that $\vartheta<\omega\tau<0$. 
This means that the electric field varies very
fast compared with the angular frequency of the oscillator.
Therefore, we may take the $e^{i\omega\tau}$ term out of the
integration in Eq. $(\rm {A1})$.
The remaining integral is easily solvable, and we get
$$I_{c}=\int_{\vartheta\omega^{-1}}^{\tau}\frac{1}{\kappa_{-}\tau '}
\frac{1}
{\left(\ln |\kappa_{-}\tau '|+\frac{1}{2}\kappa_{-}u_{0}\right)^{2l+3}}
\,d\tau ' . \eqno{\rm (A21)}$$
An explicit integration yields, then
\begin{eqnarray*}
I_{c}=&-&\frac{1}{(2l+2)\kappa_{-}}\left\{ \frac{1}{\left(\ln |\kappa_{-}
\tau |+\frac{1}{2}\kappa_{-}u_{0}\right)^{2l+2}}\right.\\ &-& 
\left.\frac{1}{\left(\ln |\kappa_{-}\vartheta\omega^{-1}|+\frac{1}{2}
\kappa_{-}u_{0}\right) ^{2l+2}} \right\} ,
\end{eqnarray*}
which for $\tau =0$ becomes
$$I_{c}(\tau =0)=\frac{1}{(2l+2)\kappa_{-}}\left(\frac{1}{2}\kappa_{-}u_{0}
\right)^{-(2l+2)}. \eqno{\rm (A22)}$$
Let us now evaluate the error of our calculation for region $c$. The exact
integral for $\tau =0$ is
\begin{eqnarray*}
I_{c}^{\rm exact}= \int_{\tau_{b}}^{0} e^{i\omega\tau '}
\frac{1}{\kappa_{-}\tau '}\left(\ln |\kappa_{-}\tau '|+\frac{1}{2}
\kappa_{-}u_{0}\right)^{-(2l+3)}\,d\tau '
\\
\approx   \int_{\tau_{b}}^{0}(1+i\omega\tau ')
\frac{1}{\kappa_{-}\tau '}\left(\ln |\kappa_{-}\tau '|+\frac{1}{2}
\kappa_{-}u_{0}\right)^{-(2l+3)}\,d\tau '
\\
=\int_{\tau_{b}}^{\tau =0}\frac{1}{\kappa_{-}\tau '}\left(\ln |\kappa_{-}
\tau '|+\frac{1}{2}\kappa_{-}u_{0}\right)^{-(2l+3)}\,d\tau '\\
+i\frac{\omega}{\kappa_{-}}\int_{\tau_{b}}^{\tau =0}
\left(\ln |\kappa_{-}\tau '|+\frac{1}{2}\kappa_{-}u_{0}\right)^{-(2l+3)}
\,d\tau '.
\end{eqnarray*} 
Now, the first term in the right-hand side of the last expression is the
same as our approximation for $I_{c}$ [see Eq. (A21)]. Therefore, we obtain
for the relative error of our approximation for region $c$:
\begin{eqnarray*}
\left|\frac{I_{c}^{\rm exact}-I_{c}^{\rm approx.}}{I_{c}^{\rm approx.}}
\right|(\tau =0)\\
\approx  \left|
\frac{\frac{\omega}{\kappa_{-}}\int_{\tau_{b}}^{\tau =0}
\left(\ln |\kappa_{-}\tau '|+\frac{1}{2}\kappa_{-}u_{0}\right)^{-(2l+3)}
\,d\tau '}{\frac{1}{(2l+2)\kappa_{-}}\left(\frac{1}{2}\kappa_{-}u_{0}
\right)^{-(2l+2)}}\right| \\
\le \left| \frac{\frac{\omega\tau_{b}}{\kappa_{-}}\left(\frac{1}{2}
\kappa_{-}u_{0}\right)^{-(2l+3)}}{\frac{1}{(2l+2)\kappa_{-}}
\left(\frac{1}{2}\kappa_{-}u_{0}\right)^{-(2l+2)}}\right|\\
\approx \frac{\omega\tau_{b}}{2l+2}\left(\frac{1}{2}
\kappa_{-}u_{0}\right)^{-1}
\approx \frac{2l+3}{2l+2}\left(\frac{1}{2}\kappa_{-}u_{0}\right)^{-2}.
\end{eqnarray*}
For the typical parameters we chose we thus obtain
$$
\left|\frac{I_{c}^{\rm exact}-I_{c}^{\rm approx.}}{I_{c}^{\rm approx.}}
\right|(\tau =0) \ll 1. \eqno{\rm (A23)} 
$$
For the evaluation of Eq. (A23) we again used the results of Ref. 
\cite{thesis}.
From the evaluation of the error associated with our approximation for the
contribution of region $c$ to $I$, it is clear that the most dominant error
in $I_{c}$ is imaginary. Therefore, we should also verify, that this
imaginary error is negligible in comparison with the imaginary part of
the contribution of region $b$ to $I$. 
Repeating the relative error
evaluation, we now obtain,
\begin{eqnarray*}
\left|\frac{I_{c}^{{\rm exact}}-I_{c}^{{\rm approx.}}}
{(E_{0}/\kappa_{-}){\rm Im}(I_{b}^{{\rm approx.}})}\right|(\tau=0)\\ 
\approx
\left|\frac{\frac{\omega}{\kappa_{-}}\int_{\tau_{b}}^{\tau =0}
\left(\ln |\kappa_{-}\tau '|+\frac{1}{2}\kappa_{-}u_{0}\right)^{-(2l+3)}
\,d\tau '}{\frac{1}{\kappa_{-}}\frac{\pi}{2}\left(
\frac{1}{2}\kappa_{-}u_{0}\right)^{-(2l+3)}}\right|\\
\le 
\left|\frac{\omega\tau_{b}}{\pi /2}\right| 
\approx
\left| \frac{2(2l+3)}{\pi}\left(\frac{1}{2}\kappa_{-}u_{0}\right)^{-1}
\right|\ll 1.
\end{eqnarray*} 

Collecting the contributions to $I$ from all three regions, we find that
the dominant contribution (at $\tau =0$) to ${\rm Re}(I)$ comes from
region $c$. This contribution is proportional to $(\kappa_{-}u_{0})^
{-(2l+2)}$. The dominant contribution (at $\tau =0$) to
${\rm Im}(I)$, however, comes from region $b$, and is proportional to
$(\kappa_{-}u_{0})^{-(2l+3)}$.
Hence, we find for the required integral
$$I(\tau =0)={\rm Re}\left[I(0)\right]+i{\rm Im}\left[I(0)\right],
\eqno{\rm (A24)}$$
where
$${\rm Re}\left[I(0)\right]
=-\frac{C'}{(2l+2)\kappa_{-}}\left(\frac{1}{2}\kappa_{-}u_{0}
\right)^{-(2l+2)}$$
$$
+O\left[\left(\kappa_{-}u_{0}\right)^{-(2l+3)}\right]
\eqno{\rm (A25)}$$
and
$${\rm Im}\left[I(0)\right]
=\frac{C'}{\kappa_{-}}\frac{\pi}{2}\left(\frac{1}{2}
\kappa_{-}u_{0}\right)^{-(2l+3)}+
O\left[\left(\kappa_{-}u_{0}\right)^{-(2l+4)}\right].
\eqno{\rm (A26)}$$
When $\tau =0$ we thus obtain, to the leading order in 
$(\kappa_{-}u_{0})^{-1}$,
$$
I(0)\approx -\frac{C'}{(2l+2)\kappa_{-}}\left(\frac{1}{2}\kappa_{-}u_{0}
\right)^{-(2l+2)}$$ 
$$
+\frac{C'}{\kappa_{-}}\frac{\pi}{2} i
\left(\frac{1}{2}\kappa_{-}u_{0}\right)^{-(2l+3)}. \eqno{\rm (A27)}
$$
Note, that $X(\tau=0)$ is dominated by region $b$
[Eq. (\ref{X0})], and $\dot{X}(\tau=0)$ and ${\cal E}_{\rm c}(\tau=0)$ are
dominated by region $c$ [Eqs. (\ref{E0}) and (\ref{V0})], as well as
${\cal E}_{\rm q}(0)$ [Eq. (\ref{Eq0})].

\section*{Appendix B: The Quantum Formalism}
Consider an electric field $E_{x}({\bar v})$ in flat spacetime,
which vanishes for
${\bar v}< {\bar v}_{0}$. After ${\bar v}={\bar v}_{0}$ the field
$E_{x}({\bar v})$ rises, and vanishes again for ${\bar v}\ge {\bar v}_{1}$.
This electric field can be obtained from a potential, whose only
non-vanishing component is $A_{x}({\bar v})$, which vanishes
for ${\bar v}< {\bar v}_{0}$, starts changing at  ${\bar v}={\bar v}_{0}$,
and assumes a final constant value $A_{0}$ for ${\bar v}\ge {\bar v}_{1}$.
(We note that any such field is consistent with the Maxwell field
equations.) We  notice that
$$
A_{0}=\int_{{\bar v}_{0}}^{{\bar v}_{1}}E_{x}({\bar v})\,d{\bar v}.
\eqno{\rm (B1)}
$$
Let us take a quantum system, which initially is in the eigenstate $|i
\rangle $. We calculate $a_{f}$ according to first-order time-dependent
perturbation theory. In the Coulomb gauge, in which our vector potential is
written, the interaction term in the Hamiltonian is $-\frac{i}{\hbar}
\frac{e}{m}{\bf A\cdot p}$. We calculate $a_{f}$ for ${\bar v}>
{\bar v}_{1}$. More exactly, we assume that the system is centered about 
${\bar z}=0$, and that its typical spatial extension is $\delta
{\bar z}$. We calculate $a_{f}$ at the time ${\bar v}_{2}>\tau_{1}+
\delta {\bar z}$, such that for all points of the system
${\bar v}>{\bar v}_{1}$. This means 
that the system has fully crossed the
region of non-vanishing electromagnetic field.
Hence, the electromagnetic field vanishes everywhere at the
instant we calculate $a_{f}$. We now notice that at that instant
$A_{x}=A_{0}\ne 0$ in
general. Therefore, the interaction term does not vanish formally, and from
first-order perturbation theory we get
$$
\dot{a}_{f}=-\frac{i}{\hbar}\frac{e}{m}
\langle f|A_{0}p_{x}|i\rangle e^{i\Omega_{fi}\tau}, \eqno{\rm (B2)}
$$
which generally is non-vanishing. Integration of (B2) yields
$$ 
a_{f}=-\frac{i}{\hbar}\frac{e}{m}\left({\rm const}+\frac{A_{0}}
{i\Omega_{fi}}
\langle f|p_{x}|i\rangle e^{i\Omega a_{fi}\tau}\right). \eqno{\rm (B3)}$$
($A_{0}$ is constant, and therefore we could take it out of the matrix
element.) This may lead us to a weird (and wrong) conclusion: even though
there is no electromagnetic field, $a_{f}$ apparently continues to evolve.
The resolution of this apparent ``paradox'' is as follows. If we write the
state of the system for ${\bar v}>{\bar v}_{2}$ in the standard gauge
appropriate for that state (namely, $A_{x}=0$ instead of $A_{x}=A_{0}$), we
find that $a_{f}$ is constant. [Initially,
since we started with $E=0$, i.e., with a vanishing perturbing field, we
chose the standard gauge for the description of the system, namely, we
chose $A_{x}=0$. (This is, of course, the gauge in which the system's
wave-functions take their standard form.) Now, when we come to interpret
the final state (in which again $E=0$), we again use the gauge which is
most natural for this situation, namely, we again take $A_{x}=0$. This
requires us to transform our original results [Eq. (\ref{QE1})] 
to the new gauge.] 

Let us show this now:
The full wave-function is given by
$$
\psi (\tau)=a_{i}(\tau)|i\rangle e^{-i\omega_{i}\tau} 
+\sum_{\{|f\rangle\}}
a_{f}(\tau)|f\rangle e^{-i\omega_{f}\tau}. \eqno{\rm (B4)}$$
We take $a_{i}(\tau)\approx 1$, which is consistent with small
transition amplitudes. Substituting (B3) in (B4) we find
(for ${\bar v}>{\bar v}_{2}$):
$$
\psi (\tau)=\left( |i\rangle-\sum_{\{|f\rangle\}}\frac{i}{\hbar}\frac{e}{m}
\frac{A_{0}}{i\Omega_{fi}}\langle f|p_{x}|i\rangle |f\rangle\right)
e^{-i\omega_{i}\tau}$$ 
$$-\sum_{\{|f\rangle\}}
{\rm c}_{f}\frac{1}{\hbar}\frac{e}{m}|f\rangle
e^{-i\omega_{f}\tau}. \eqno{\rm (B5)}$$ 
The constant term in Eq. (B3) may depend on $f$, and therefore was given in 
Eq. (B5) an appropriate index. The gauge transformation 
$$A_{x}^{\rm new}=A_{x}^{\rm old}-A_{0}  \eqno{\rm (B6)}$$ 
of the 
potential leads to the following gauge transformation of the wave-function: 
$$\psi^{\rm new}={\rm exp}\left(
\frac{ie}{\hbar}\int A_{0}\,dx\right)\psi^{\rm old}.
\eqno{\rm (B7)}$$
Since we work to first order in the perturbation, we may use 
$$\psi^{\rm new}\approx \left( 1+\frac{ie}{\hbar}A_{0}x\right) 
\psi^{\rm old}. 
\eqno{\rm (B8)}$$
Substituting Eq. (B5) for $\psi^{\rm old}$ we get that 
\begin{eqnarray}
\psi^{\rm new}&\approx&\left( 1+\frac{ie}{\hbar}A_{0}x\right)
\left[\left(|i\rangle-\sum_{\{|f\rangle\}}
\frac{i}{\hbar}\frac{e}{m}\frac{A_{0}}{i\Omega_{fi}}
\langle f|p_{x}|i\rangle |f\rangle
\right)\right. \nonumber\\ 
&\times & \left.e^{-i\omega_{i}\tau} 
- \sum_{\{|f\rangle\}}{\rm c}_{f}\frac{i}{\hbar}
\frac{e}{m}|f\rangle e^{-i\omega_{f}\tau}\right]\nonumber \\
&\approx&\left(|i\rangle+\frac{ie}{\hbar}A_{0}x|i\rangle-
\sum_{\{|f\rangle\}}\frac{1}{\hbar}
\frac{e}{m}\frac{A_{0}}{\Omega_{fi}}
\langle f|p_{x}|i\rangle|f\rangle\right)\nonumber\\
&\times & e^{-i\omega_{i}\tau} 
- \sum_{\{|f\rangle\}}{\rm c}_{f}\frac{i}{\hbar}\frac{e}{m}|f\rangle
e^{-i\omega_{f}\tau}\nonumber 
\end{eqnarray} 
and therefore we obtain
$$\psi^{\rm new}\approx |i\rangle e^{-i\omega_{i}\tau}-
\sum_{\{|f\rangle\}}{\rm c}_{f}
\frac{i}{\hbar}\frac{e}{m}|f\rangle e^{-i\omega_{f}\tau}, \eqno{\rm (B9)}$$
after keeping linear terms in the interaction only, and using the
relation
$$\sum_{\{|f\rangle\}}\frac{1}{\Omega_{fi}}
\langle f|p_{x}|i\rangle |f\rangle =imx|i\rangle.
\eqno{\rm (B10)}$$
Hence, we find that
the transformation to the natural gauge $(A=0)$
removes the second term in the
brackets in Eq. (B5). We now get the following conclusion: The correct way
to understand the result of the perturbative calculation is in the gauge
where the vector potential vanishes. In this gauge, the second term in Eq.
(B3) vanishes, and we obtain that
$$a_{f}^{\rm new}={\rm const}, \eqno{\rm (B11)}$$
as should be expected (because the electromagnetic field vanishes).

Finally, we find the specific form for the constant term of Eqs. (B3) and
(B11). To obtain this, we 
compare $a_{fi}^{(2)}$ [see Eq. (\ref{Q3})] to the right-hand side of Eq. 
(B3) and obtain that
$$a_{f}^{\rm final, new}=
-\frac{e}{m}\frac{1}{\hbar\Omega_{fi}}\int_{-\infty}^{\tau}\langle 
\psi_{f}|
{\bf E\cdot p}|\psi_{i}\rangle e^{i\Omega_{fi}t}\,dt . \eqno{\rm (B12)}$$
After the perturbation vanishes, this expression is constant.

\end{document}